\documentclass[aps,pra,nobalancelastpage,superscriptaddress, twocolumn,10pt]{revtex4-2}

\usepackage{color,ifthen,amsthm,amsmath,amsxtra,amsfonts,dsfont,graphicx,bm,tikz,scalerel,wasysym,bbm,graphicx,amsthm,braket,physics,empheq,CircuitTikz,float}
\usepackage[colorlinks=true,linkcolor=blue, citecolor=blue, urlcolor=blue, bookmarks]{hyperref}

\newcommand{\be}{\begin{equation}}
\newcommand{\ee}{\end{equation}}
\newcommand{\bea}{\begin{eqnarray}}
\newcommand{\eea}{\end{eqnarray}}
\newcommand{\bse}{\begin{subequations}}
\newcommand{\ese}{\end{subequations}}

\theoremstyle{plain}
\newcommand{\1}{\mathbbm{1}}

\theoremstyle{plain}

\theoremstyle{plain}

\begin{document}
\title{Entanglement in dual unitary quantum circuits with impurities}

\author{Shachar Fraenkel}
\affiliation{Raymond and Beverly Sackler School of Physics and Astronomy, Tel Aviv
University, Tel Aviv 6997801, Israel}

\author{Colin Rylands}
\affiliation{Centre for Fluid and Complex Systems, Coventry University, Coventry, CV1 2TT, United Kingdom}
\affiliation{SISSA and INFN Sezione di Trieste, via Bonomea 265, 34136 Trieste, Italy}

\begin{abstract}Bipartite entanglement entropy is one of the most useful characterizations of universal properties in a many-body quantum system.  Far from equilibrium, there exist two highly effective theories describing its dynamics -- the quasiparticle and membrane pictures. In this work we investigate entanglement dynamics, and these two complementary approaches, in a quantum circuit model perturbed by an impurity. In particular, we consider a dual unitary quantum circuit containing a spatially fixed, non-dual-unitary impurity gate, allowing for differing local Hilbert space dimensions to either side.  We compute the entanglement entropy for both a semi-infinite and a finite subsystem within a finite distance of the impurity, comparing exact results to predictions of the effective theories. We find that for a semi-infinite subsystem,
both theories agree with each other and the exact calculation. For a finite subsystem, 
however, both theories qualitatively differ, with the quasiparticle picture predicting a non-monotonic growth in contrast to the membrane picture. We show that such non-monotonic behavior can arise even in random chaotic circuits, shedding light on the range of validity of the membrane picture in such systems.
\end{abstract}

\maketitle

\textit{Introduction}---
One dimensional quantum systems are particularly susceptible to the influence of interactions~\cite{giamarchi2004quantum}.  A prime example of this is found by coupling such a system to a quantum impurity.  This can have a drastic effect not only on the transport of a system~\cite{kane1992transport} but also on its spectral properties.  Indeed a single impurity added to an otherwise integrable system can be enough for it to exhibit properties of quantum chaos~\cite{santos2004integrability,barisic2009incoherent,santos2014local,znidaric2020weak,pandey2020adiabatic,brenes2020eigenstate,brenes2020low,fritzsch2022boundary,fritzsch2023boundary}.
Despite this, however,  the underlying integrability of the pure system can manifest itself in certain settings such as ballistic transport of charge in a non-equilibrium steady state~\cite{brenes2018high}.  Quantum impurities thus provide an interesting probe with which to study the behavior of many-body systems, and the interplay with integrability breaking, particularly out of equilibrium.  

For a non-equilibrium system, one of the most fundamental and insightful quantities one can calculate is the bipartite entanglement entropy ~\cite{amico2008entanglement}.  Two complementary effective theories exist to describe this quantity: the quasiparticle picture~\cite{calabrese2005evolution} and the entanglement membrane picture~\cite{jonay2018coarse,nahum2020entanglement}.  In the former,  entanglement between spatial regions is created by the propagation of correlated sets of quasiparticles~\cite{calabrese2005evolution}.  By calculating how these quasiparticles are correlated initially, one can make quantitative predictions on the growth of entanglement. The theory applies to integrable models, free or interacting~\cite{alba2017entanglement}, and has been adapted also to many other scenarios~\cite{calabrese2012quantum,coser2014entanglement,alba2019quantum,horvath2024full,bertini2023nonequilibrium,parez2021quasiparticle,parez2021exact,
dubail2017entanglement,parez2022analytic,rath2023entanglement,ares2023entanglement,rylands2024microscopic,bertini2024dynamics,rottoli2024entanglement}.  In the latter theory, 
entanglement does not spread throughout the system via quasiparticles but is instead produced locally and can be computed by studying a membrane in space-time which separates the subsystem from its complement~\cite{jonay2018coarse,nahum2020entanglement}.  The only input this requires is the energy or tension associated to the membrane, however this is difficult to determine and has only been achieved in a limited number of cases~\cite{nahum2020entanglement,rampp2024entanglement,foligno2024quantum}.  In contrast to the quasiparticle picture, the membrane theory applies to chaotic models. For contiguous subsystems, both theories provide the same qualitative results, namely the linear growth of entanglement from lowly entangled states,  a fact which can be understood using the unifying approach of space-time duality~\cite{bertini2022growth}.  For finite-size systems or disjoint subsystems,  however, the two approaches give qualitatively different predictions~\cite{nahum2017quantum,foligno2024entanglement,alba2019scrambling,modak2020entanglement} revealing the distinct properties of the models to which they apply.

\begin{figure}
  \includegraphics[viewport=20bp 200bp 560bp 600bp,clip,width=1\columnwidth]{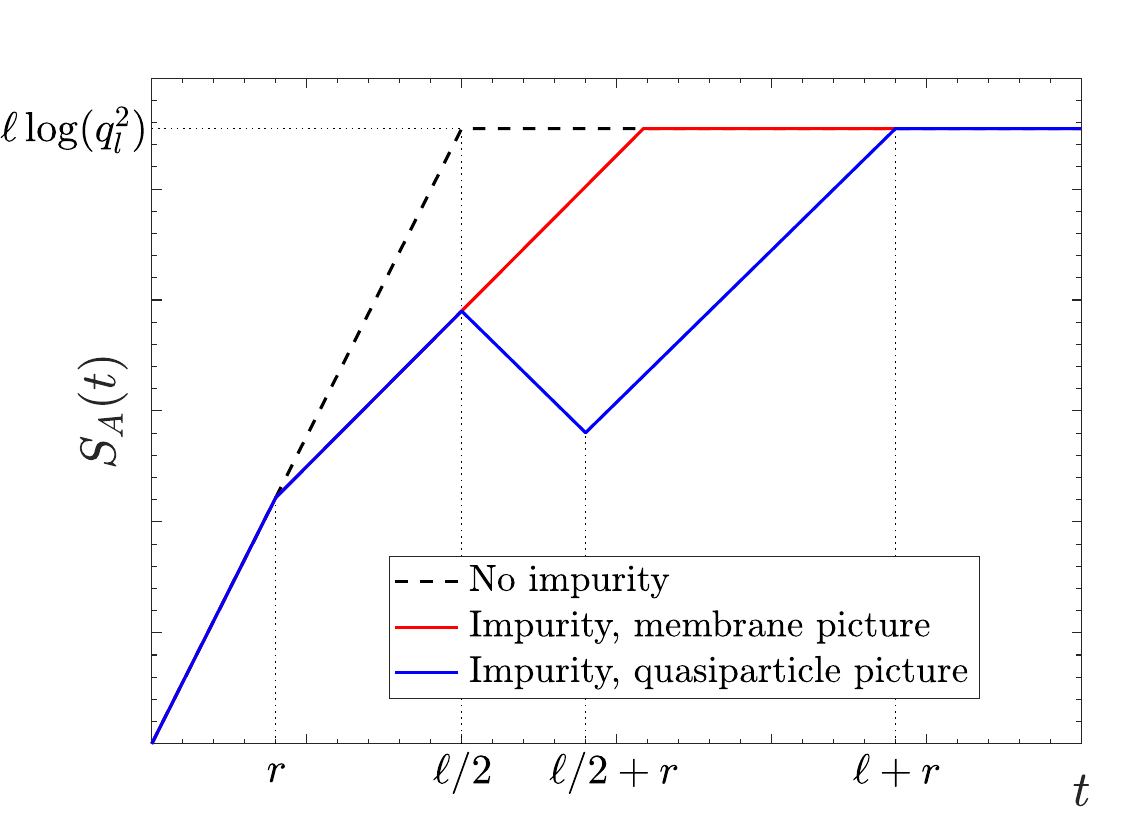}
  \caption{\label{fig:ent} Entanglement entropy dynamics for a subsystem $A$ with length $2\ell$ in a dual unitary circuit (black dashed line), and in a dual unitary circuit perturbed by an impurity (solid lines). In circuits with an impurity, $A$ is at a distance $2r$ from it. The blue line is for bulk swap gates with local Hilbert space dimension $q_l=q_r=2$ and the impurity given in ~\eqref{eq:Imp_example}, with $c=0.05$, which matches the quasiparticle prediction.  The red line represents the prediction of the membrane picture for the same impurity gate but with maximally chaotic dual unitary bulk gates. }
\end{figure}

An ideal setting for the study of these contrasting approaches is given by dual unitary quantum circuits. These constitute a particular class of many-body quantum systems that describes a wide range of dynamics from integrable to chaotic while at the same time retaining a degree of solubility~\cite{bertini2018exact,gopalakrishnan2019unitary,prosen2021many,claeys2021ergodic,kos2021correlations,kos2023circuits,foligno2023temporal,piroli2020scrambling,bertini2020operator,bertini2020operator2,holdendye2023fundamental,borsi2022construction}. Many of their properties, including their bipartite entanglement entropy, have been calculated exactly~\cite{bertini2019exact,piroli2020exact,giudice2022temporal,vernier2023integrable}, and recently they also provided the first analytical verification of the discrepancy between the membrane and quasiparticle pictures in the case of disjoint intervals~\cite{foligno2024entanglement}. Moreover, they can be simulated on modern quantum computing platforms, see e.g. \cite{bertini2025exactly}.

In this work we consider a hybrid system consisting of two dual unitary circuits, with possibly different local Hilbert space dimensions,  stitched together via a quantum impurity.  The impurity is represented by a line of gates at a fixed position which are not dual unitary.  In this setup we calculate the entanglement entropy between a subsystem that does not contain the impurity and its complement, for both
semi-infinite and finite subsystems, and compare with the predictions of the quasiparticle and membrane pictures. We find that for a semi-infinite subsystem, the exact calculation can be interpreted using either the quasiparticle or membrane pictures.

For a finite subsystem, however, the two predictions qualitatively differ, see Fig.~\ref{fig:ent}.  In particular, after an initial period of linearly growing entanglement, the quasiparticle picture predicts a period of decreasing entanglement entropy followed by a return to growth and an eventual saturation.  This is in contrast to the membrane picture,  which always predicts monotonic growth before saturation. Indeed, our exact calculation agrees with the quasiparticle picture for an integrable circuit, and with the membrane picture for a maximally-ergodic random circuit. Surprisingly, we find that, in between these two extremes, the non-monotonic behavior appears even for chaotic random circuits, so that the membrane picture cannot fully describe their entanglement dynamics.

\textit{Setting}---
We consider a one dimensional chain of $2L$ qudits with $L$ even, located at coordinates $x=1,\dots, 2L$.  Each qudit has local Hilbert space dimension $q_x$ where $q_{x\leq L}=q_l$ and $q_{x> L}=q_r$.  The system is initialized in the state
\be \label{eq:initial_state}
\ket{\Psi(0)}=\bigotimes_{x=1}^L\ket{m_x},\,\, 
\ket{m_x} \!=\frac{1}{q_{2x}^{1/2}}\!\sum_{i,j=1}^{q_{2x}}[m_x]_{ij}\ket{ij}_x,
\ee
where $\ket{ij}_x=\ket{i}_{2x-1}\otimes\ket{j}_{2x}$ with $\ket{i}_y,i=1,\dots,q_y$ spanning the set of states of the qudit at site $y$, and $\{m_x\}$ is a set of $q_{2x}\times q_{2x}$ unitary matrices characterizing the initial state.  Such states are the simplest type of solvable initial states for a dual unitary circuit~\cite{piroli2020exact}. The dynamics of the system are generated by a brickwork-pattern quantum circuit, such that the state of the system after $t+1$ time steps is given by 
\be\label{eq:time_evo_op}
\ket{\Psi(t+1)}=\mathbb{U}(t+1)\ket{\Psi(t)},
\ee
and where the time-evolution operator is 
\be 
 \mathbb{U}(t)=
\smashoperator{\bigotimes_{x\,{\rm odd}}} U_{x,x+1}(x,t)
\smashoperator{\bigotimes_{x'\,{\rm even}}} U_{x'\!\!,x'+1}(x',t).
\ee
Here $U_{x,x+1}(x,t)$ are unitary operators acting on the whole chain, but non-trivially only at sites $x,x+1$ as $U(x,t)$,  and as the identity elsewhere.  The $q_xq_{x+1}\times q_xq_{x+1}$ matrices $U(x,t)$ are called local gates, and in our system we take them to be of one of two different types, a bulk gate or an impurity gate.  The bulk gates comprise almost all of the system except for a line of impurity gates located at $x=L,~\forall t$.  
While both types of gates are unitary when viewed in the time-like direction, i.e.~$U(x,t)U^\dag(x,t)=\1$, $\forall x,t$,  the bulk gate is also dual unitary, meaning it is unitary in the space-like direction as well.  The local gates thus obey
$\tilde{U}(x,t)\tilde{U}^\dag(x,t)=\1,~\forall x\neq L$, where 
\begin{eqnarray}
~\!\bra{ij}\tilde{U}(x,t)\ket{kl}=~\!\bra{jl}U(x,t)\ket{ik}.
\end{eqnarray}
We need not take any of the bulk or impurity gates to be the same, and indeed later on we will examine a system of randomly chosen bulk gates.  
When $q_l=q_r$, it is possible to take the limit where the impurity gate is also dual unitary and reproduce results for the pure dual unitary circuit.  When $q_l\neq q_r$, the impurity can never be made dual unitary. 

Our quantity of interest is the R\'enyi entanglement entropy between a subsystem $A$ and its complement $\bar{A}$,
\be
S^{(n)}_A(t)=\frac{1}{1-n}\log \tr [\rho^n_A(t)],
\ee
where $\rho_A(t)=\tr_{\bar A}[\ketbra{\Psi(t)}{\Psi(t)}]$ is the reduced density matrix of $A$. Taking the replica limit, $n\to1 $,  we obtain the von Neumann entanglement entropy $S_A(t)=\lim_{n\to 1}S^{(n)}_A(t)$. The subsystem is composed of  $2\ell$ contiguous sites a distance $2r$ from the impurity, $A= (L-2\ell-2r,L-2r]$.

We may represent our quantum circuit and conveniently perform calculations  diagrammatically using the standard graphical notation of tensor networks~\cite{cirac2021matrix}. Specifically, we represent the folded, replicated local gate as
\bea
\left[U(x,t)\otimes U^{\star}(x,t)\right]^{\otimes_{\rm{r}} n}&=&
  \begin{tikzpicture}[baseline={([yshift=-0.6ex]current bounding box.center)},scale=0.5]
   \prop{0}{0}{colSt}{}
 \end{tikzpicture}~{\rm or} ~ \begin{tikzpicture}[baseline={([yshift=-0.6ex]current bounding box.center)},scale=0.5]
   \newprop{0}{0}{colSt}{}
 \end{tikzpicture},\\
 \left[U(L,t)\otimes U^{\star}(L,t)\right]^{\otimes_{\rm{r}} n}&=&
  \begin{tikzpicture}[baseline={([yshift=-0.6ex]current bounding box.center)},scale=0.5]
   \propimp{0}{0}{colimp}{}
 \end{tikzpicture},
\eea
where we have denoted by $\otimes_{\rm{r}}$ the tensor product over different replicas. When necessary, we shall distinguish between different local Hilbert space dimensions using thin or thick lines for $q_l,q_r$, respectively. 
We also introduce the two special states on the replicated Hilbert space and their corresponding graphical depictions
\begin{equation}
  \begin{aligned}
&  \smashoperator{\sum_{s_{j}\in 
\mathbb{Z}_{q_x}}}\ket{s_1,s_1,s_2,s_2,\dots ,s_n,s_n}=
 \begin{tikzpicture}[baseline={([yshift=-0.6ex]current bounding box.center)},scale=0.5]
   \gridLine{0}{0}{0}{0.75}
   \circle{0}{0}
 \end{tikzpicture}~,\\ 
 & \smashoperator{\sum_{s_{j}\in \mathbb{Z}_{q_x}}}\ket{s_1,s_2,s_2,s_3,\dots,s_n,s_1}=
 \begin{tikzpicture}[baseline={([yshift=-0.6ex]current bounding box.center)},scale=0.5]
   \gridLine{0}{0}{0}{0.75}
   \square{0}{0}
 \end{tikzpicture}~.
  \end{aligned}
\end{equation}
The first state, represented by the empty circle, couples together the forward and backward time evolution on the same replica, whereas the second state instead couples the backward time evolution in one copy to the forward time evolution in the next copy.  

The unitarity of the local gates is expressed graphically as 
\be
\begin{aligned}
\label{eq:local_gate_rels_1}
 \begin{tikzpicture}[baseline={([yshift=-0.6ex]current bounding box.center)},scale=0.5]
   \prop{0}{0}{colSt}{}
   \circle{0.5}{0.5}
    \circle{-0.5}{0.5}
 \end{tikzpicture}=
\begin{tikzpicture}[baseline={([yshift=-0.6ex]current bounding box.center)},scale=0.5]
    \gridLine{0}{0}{0}{0.75}
   \circle{0}{0.75}
    \gridLine{.5}{0}{.5}{0.75}
   \circle{.5}{0.75}
 \end{tikzpicture}~,
 \begin{tikzpicture}[baseline={([yshift=-0.6ex]current bounding box.center)},scale=0.5]
   \prop{0}{0}{colSt}{}
   \square{0.5}{0.5}
    \square{-0.5}{0.5}
 \end{tikzpicture}=
 \begin{tikzpicture}[baseline={([yshift=-0.6ex]current bounding box.center)},scale=0.5]
    \gridLine{0}{0}{0}{0.75}
   \square{0}{0.75}
    \gridLine{.5}{0}{.5}{0.75}
   \square{.5}{0.75}
 \end{tikzpicture},
  \begin{tikzpicture}[baseline={([yshift=-0.6ex]current bounding box.center)},scale=0.5]
   \propimp{0}{0}{colimp}{}
   \circle{0.5}{0.5}
    \circle{-0.5}{0.5}
 \end{tikzpicture}=
\begin{tikzpicture}[baseline={([yshift=-0.6ex]current bounding box.center)},scale=0.5]
    \gridLine{0}{0}{0}{0.75}
   \circle{0}{0.75}
    \newnctgridLine{.5}{0}{.5}{0.75}
   \circle{.5}{0.75}
 \end{tikzpicture}~,
 \begin{tikzpicture}[baseline={([yshift=-0.6ex]current bounding box.center)},scale=0.5]
   \propimp{0}{0}{colimp}{}
   \square{0.5}{0.5}
    \square{-0.5}{0.5}
 \end{tikzpicture}=
 \begin{tikzpicture}[baseline={([yshift=-0.6ex]current bounding box.center)},scale=0.5]
    \gridLine{0}{0}{0}{0.75}
   \square{0}{0.75}
    \newnctgridLine{.5}{0}{.5}{0.75}
   \square{.5}{0.75}
 \end{tikzpicture},
\end{aligned} 
 \ee
 along with the same diagrams flipped about the horizontal axis.  For the bulk gates, we have the additional dual unitarity conditions
 \be
\begin{aligned}
\label{eq:local_gate_rels_2}
 &\begin{tikzpicture}[baseline={([yshift=-0.6ex]current bounding box.center)},scale=0.5]
   \prop{0}{0}{colSt}{}
   \circle{-0.5}{-0.5}
    \circle{-0.5}{0.5}
 \end{tikzpicture}=
\begin{tikzpicture}[baseline={([yshift=-0.6ex]current bounding box.center)},scale=0.5]
  \circle{-0.5}{-0.5}
    \circle{-0.5}{0.5}
    \gridLine{-0.38}{-0.5}{0.3}{-0.5}
      \gridLine{-0.38}{0.5}{0.3}{0.5}
 \end{tikzpicture}~,& &
 \begin{tikzpicture}[baseline={([yshift=-0.6ex]current bounding box.center)},scale=0.5]
   \prop{0}{0}{colSt}{}
   \square{-0.5}{-0.5}
    \square{-0.5}{0.5}
 \end{tikzpicture}=
 \begin{tikzpicture}[baseline={([yshift=-0.6ex]current bounding box.center)},scale=0.5]
      \square{-0.5}{-0.5}
    \square{-0.5}{0.5}
        \gridLine{-0.38}{-0.5}{0.3}{-0.5}
      \gridLine{-0.38}{0.5}{0.3}{0.5}
 \end{tikzpicture}~,
 \end{aligned}
 \ee
 along with the same diagrams flipped about the vertical axis. The replicated initial state is represented as
 \begin{align}
&\left(\ket{m_x}\otimes\ket{m_{x}}^\star\right)^{\otimes_{\rm r}n}=\begin{tikzpicture}[baseline={([yshift=-0.6ex]current bounding box.center)},scale=0.65]
  \tgridLine{0}{0}{-0.25}{0.25}
  \tgridLine{1}{0}{1.25}{0.25}
  \istate{0}{0}{colSt}
\end{tikzpicture}~,
 \end{align}
 which due to the unitarity of the matrix $m_x$ obeys the following relations:
 \begin{eqnarray}
 \begin{tikzpicture}[baseline={([yshift=-0.6ex]current bounding box.center)},scale=0.65]
 \tgridLine{1}{0}{1.25}{0.25}
  \istate{0}{0}{colSt}
    \circle{-0.12}{0.12}
\end{tikzpicture}=\begin{tikzpicture}[baseline={([yshift=-0.6ex]current bounding box.center)},scale=0.65]
  \tgridLine{0}{0}{0.5}{0.5}
  \circle{0}{0}
   \circle{-1.3}{-.5} 
     \tgridLine{-1.15}{-.5}{-0.8}{-0.5}
     \circle{-.8}{-.5};
     \draw[semithick](-1.45,-.1)--(-.6,-.1)
     node[midway,yshift=6.5pt]{$1$};
;\end{tikzpicture}~,~\begin{tikzpicture}[baseline={([yshift=-0.6ex]current bounding box.center)},scale=0.65]
  \tgridLine{1}{0}{1.25}{0.25}
  \istate{0}{0}{colSt}
    \square{-0.15}{0.15}
\end{tikzpicture}=\begin{tikzpicture}[baseline={([yshift=-0.6ex]current bounding box.center)},scale=0.65]
  \tgridLine{0}{0}{0.5}{0.5}
  \square{0}{0}
   \square{-1.3}{-.5} 
     \tgridLine{-1.15}{-.5}{-0.8}{-0.5}
     \square{-.8}{-.5};
     \draw[semithick](-1.45,-.1)--(-.6,-.1)
     node[midway,yshift=6.5pt]{$1$};
\end{tikzpicture}~,
 \end{eqnarray}
 along with the same diagrams flipped about the vertical axis.  Here the coefficients $\begin{tikzpicture}[baseline={([yshift=-0.6ex]current bounding box.center)},scale=0.65]
  \circle{-1.3}{-.5} 
     \tgridLine{-1.15}{-.5}{-0.8}{-0.5}
     \circle{-.8}{-.5};
\end{tikzpicture}=\begin{tikzpicture}[baseline={([yshift=-0.6ex]current bounding box.center)},scale=0.65]
  \square{-1.3}{-.5} 
     \tgridLine{-1.15}{-.5}{-0.8}{-0.5}
     \square{-.8}{-.5};
\end{tikzpicture}=\left(q_{2x}\right)^n$ come from the normalization of the initial state. Using this notation we can depict $\tr[\rho^n_A(t)]$ using the diagram in Fig.~\ref{fig:rho_circuit}.

\begin{figure}
\begin{tikzpicture}[baseline={([yshift=0.6ex]current bounding box.center)},scale=0.45]
    \foreach \x in {-5,-3,...,5}{\prop{\x}{-3}{colSt}{}}
     \foreach \x in {7,9}{\newprop{\x}{-3}{colSt}{}}
      \foreach \x in {-5,-3,...,5}{\prop{\x}{-1}{colSt}{}}
           \foreach \x in {7,9}{\newprop{\x}{-1}{colSt}{}}
        \foreach \x in {-5,-3,...,5}{\prop{\x}{1}{colSt}{}}
             \foreach \x in {7,9}{\newprop{\x}{1}{colSt}{}}
          \foreach \x in {-5,-3,...,5}{\prop{\x}{3}{colSt}{}}
               \foreach \x in {7,9}{\newprop{\x}{3}{colSt}{}}
    \foreach \x in {-6,-4,...,6}{\prop{\x}{-4}{colSt}{}}
         \foreach \x in {8,10}{\newprop{\x}{-4}{colSt}{}}
     \foreach \x in {-6,-4,...,6}{\prop{\x}{-2}{colSt}{}}
          \foreach \x in {8,10}{\newprop{\x}{-2}{colSt}{}}
    \foreach \x in {-6,-4,...,6}{\prop{\x}{0}{colSt}{}} 
         \foreach \x in {8,10}{\newprop{\x}{0}{colSt}{}}
     \foreach \x in {-6,-4,...,6}{\prop{\x}{2}{colSt}{}}
         \foreach \x in {8,10}{\newprop{\x}{2}{colSt}{}}
      \foreach \x in {-4,-2,0,2} {\propimp{6}{\x}{colimp}{}}
    \foreach \x in {-5.5,-3.5,...,5.5}{\istate{\x}{-4.5}{colSt}}
        \foreach \x in {6.5,8.5}{\newistate{\x}{-4.5}{colSt}}
    \foreach \x in {-4,...,1}{\square{\x+0.5}{3.5}}
     \foreach \x in {2,...,9}{\circle{\x+0.5}{3.5}}
       \foreach \x in {-6,-5}{\circle{\x+0.5}{3.5}}
 \draw[semithick,decorate,decoration={brace}] (-3.75,4) -- (1.75,4) node[midway,yshift=7.5pt] {$2\ell$};
  \draw[semithick,decorate,decoration={brace}] (2.25,4) -- (5.75,4) node[midway,yshift=7.5pt] {$2r$};
  \draw[semithick,decorate,decoration={brace}] (11,3) -- (11,-4.5) node[midway,xshift=12.5pt,rotate= 270] {$2t$};
    \draw[thick,dashed,red] (2,3) -- (5.5,-0.5) -- (5.5,-4.75) ;
     \draw[thick,dashed,red] (-3.5,3) -- (-3.5,-4.75) ;
  \end{tikzpicture}
  \caption{\label{fig:rho_circuit}The diagrammatic representation of $\tr[\rho^n_A(t)]$ with $\ell=3,r=2$ and for $t=4$. Here, time runs vertically, from the bottom of the figure to the top. Orange boxes denote bulk dual unitary operators, green represents the impurity, and the orange circles denote the initial state. The local Hilbert space dimension to the left and right of the impurity is $q_l$ and $q_r$, respectively. This is depicted using thin and thick lines.   The dashed red lines indicate the paths of the entanglement membranes before the entropy saturates. }
\end{figure}

\textit{Semi-infinite subsystem}--- We begin with the calculation of the bipartite R\'enyi entanglement entropy of a  semi-infinite subsystem, $\ell=L/2-r$ and $L\to \infty$.  Our starting point is the diagrammatic representation of $\tr[\rho^n_A(t)]$ shown in Fig.~\ref{fig:rho_circuit}.  Using the unitarity of the local gates,~\eqref{eq:local_gate_rels_1}, this can be significantly reduced so that all gates can be removed apart from those forming a backwards light cone emanating from the interface between $A$ and $\bar{A}$.  The result is that 
\begin{align}
\label{eq:D_def}
\!\!\!\tr[\rho^n_A(t)]=\hspace{-1cm} \begin{tikzpicture}[baseline={([yshift=0.6ex]current bounding box.center)},scale=0.45]
{\prop{2}{2}{colSt}{}}
  \foreach \x in {1,3}{\prop{\x}{1}{colSt}{}}   
    \foreach \x in {0,2,...,4}{\prop{\x}{0}{colSt}{}}
    \foreach \x in {-1,1,...,5}{\prop{\x}{-1}{colSt}{}}
    \foreach \x in {-2,0,...,6}{\prop{\x}{-2}{colSt}{}}
    \foreach \x in {-3,-1,...,5}{\prop{\x}{-3}{colSt}{}}
        \foreach \x in {7}{\newprop{\x}{-3}{colSt}{}}
    \foreach \x in {-4,-2,...,6}{\prop{\x}{-4}{colSt}{}}
     \foreach \x in {8}{\newprop{\x}{-4}{colSt}{}}
   \foreach \x in {-4,-2} {\propimp{6}{\x}{colimp}{}}
    \foreach \x in {-5.5,-3.5,...,4.5}{\istate{\x}{-4.5}{colSt}}
        \foreach \x in {6.5,8.5}{\newistate{\x}{-4.5}{colSt}}
    \foreach \x in {-6,...,1}{\square{\x+0.5}{\x+1.5}}
    \foreach \x in {-6,...,1}{\circle
    {4-\x-0.5}{\x+1.5}}
     \draw[semithick,decorate,decoration={brace}] (2.55,2.85) -- (10,-4.5) node[midway,xshift=7.5pt,yshift=7.5pt] {$2t$};
  \end{tikzpicture}
  \hspace{-.5cm}, 
\end{align}
where we have depicted explicitly the case of $t=4$ and $r=2$. From this we see that the impurity gate does not appear if $t\leq r$, in which case we recover the pure dual unitary result.  The diagram can likewise be reduced when $t>r$, whereupon the impurity gate affects the growth of entanglement. Relying on the property $\begin{tikzpicture}[baseline={([yshift=-0.6ex]current bounding box.center)},scale=0.65]
  \square{-1.3}{-.5} 
     \tgridLine{-1.15}{-.5}{-0.8}{-0.5}
     \circle{-.8}{-.5};
\end{tikzpicture}=\begin{tikzpicture}[baseline={([yshift=-0.6ex]current bounding box.center)},scale=0.65]
  \circle{-1.3}{-.5} 
     \tgridLine{-1.15}{-.5}{-0.8}{-0.5}
     \square{-.8}{-.5};
\end{tikzpicture}=q_x$, the final result is
\begin{eqnarray}\label{eq:semi_infinite_S}
S^{(n)}_A(t)=\begin{cases} t\log(q_l^2) &t\leq r,\\
 r\log(q_l^2)+(t-r)S^{(n)}_{\text{imp}}&t>r,
\end{cases}
\end{eqnarray}
where $S^{(n)}_{\text{imp}}$ quantifies the operator entanglement \cite{rather2020creating} of the impurity gate, and is given by 
\begin{eqnarray}\label{eq:ImpurityOE}
e^{(1-n)S^{(n)}_{\text{imp}}}=\frac{1}{q_l^{n}q_r^n}\,\begin{tikzpicture}[baseline={([yshift=-0.6ex]current bounding box.center)},scale=0.5]
   \propimp{1}{0}{colimp}{}
   \circle{1.5}{0.5}
    \square{0.5}{0.5}
       \circle{1.5}{-0.5}
    \square{0.5}{-0.5}
 \end{tikzpicture}~.
\end{eqnarray}
This satisfies $0\leq S^{(n)}_{\text{imp}}\leq {\rm min}\{\log(q_l^2),\log(q_r^2)\}$, where the lower bound is saturated by a purely reflecting impurity, i.e., $U(L,t)=\1$. The upper bound can be achieved by any gate acting as dual unitary on a ${\rm min}\{q_l^2,q_r^2\}$-dimensional subspace and as the identity on the rest. 

As an example, consider the case $q_l=q_r=q$ and $U(L,t)=\frac{\1+i cP}{1+i c}$, $c\in\mathbb{R}$, where $P$ is the swap gate. This choice of gate is dual unitary if $c\to\infty$ but otherwise is not. We find   
\begin{equation}\label{eq:Imp_example}
S^{(n)}_{\text{imp}}=\frac{1}{1-n}\log\!\left[\frac{q^2-1}{q^{2n}}\mathcal{T}^n+\left(\mathcal{R}+\frac{1}{q^2}\mathcal{T}\right)^n\right], 
\end{equation}
where $\mathcal{T}=c^2/(1+c^2),\mathcal{R}=1-\mathcal{T}$ are the transmission and reflection coefficients of the impurity~\cite{Note11}\footnotetext[11]{See the Supplemental Material, which includes Refs.~\cite{zanardi2001entanglement,rampp2023spreading}, for details on (i) Correlations across the impurity; (ii) The quasiparticle picture prediction for the entanglement entropy, broken down to contributions of quasiparticle pairs; (iii) The membrane picture prediction for entanglement dynamics;
(iv) Random gates with maximal entangling power, including the proof of~\eqref{eq:maximal_ent}; and (v) Proofs of the bounds~\eqref{eq:bound} and~\eqref{eq:second_bound}}.  In the replica limit and for large $q$, the leading order term is
\begin{eqnarray}\label{eq:Kondo_vN_ent}
S_{\text{imp}}=\mathcal{T}\log (q^2)-\mathcal{T}\log \mathcal{T}-\mathcal{R}\log\mathcal{R}~.
\end{eqnarray}
The result~\eqref{eq:semi_infinite_S} is valid for any choice of bulk and impurity gates,  integrable or chaotic.  Accordingly, it can be interpreted using either the quasiparticle or entanglement membrane pictures.  Within the quasiparticle picture, the change in entanglement growth at $t=r$ can be seen as arising from the fact that  
right-moving quasiparticles produced at points 
$L-2r<x\le L$ have had time to scatter off the impurity and enter the subsystem~\cite{Note11}. 
Prior to $t=r$ no quasiparticles have had the time to do so, 
hence the entanglement increases as in the pure case, where counter-propagating quasiparticle pairs generate maximal entanglement. In \eqref{eq:Kondo_vN_ent} we can interpret the first term as this maximal entanglement rate reduced by the transmission coefficient. The last two terms are the entanglement entropy between the transmitted and reflected parts of a single quasiparticle that is split by the impurity, as previously discovered in integrable quantum impurity models~\cite{fraenkel2021entanglement,fraenkel2023extensive,fraenkel2024extensive,fraenkel2024exact,capizzi2023domain,rylands2023transport}. Note that the quasiparticle picture still applies despite the fact that the impurity generically breaks the integrability of the system~\cite{fritzsch2022boundary,fritzsch2023boundary}.

 For chaotic models, the membrane picture is appropriate. In the pure case the tension of the membrane,  calculated via the operator entanglement~\cite{rampp2024entanglement},  is constant.  Hence for $t\leq r$ the membrane takes any directed downward path emanating from the entangling point.  When $t>r$, however, it is more energetically favorable for the membrane to travel along the impurity rather than the bulk since, not being dual unitary, it generates less entanglement.  As a result,  the optimal path consists of the membrane traveling to the impurity along the light cone until it reaches $x=L$ and then proceeding vertically down the impurity, see red dashed lines in Fig.~\ref{fig:rho_circuit}.  Inserting a projector onto the state $\begin{tikzpicture}[baseline={([yshift=-0.6ex]current bounding box.center)},scale=0.5]
    \square{-0.5}{0.5}
      \gridLine{-0.38}{0.5}{0.2}{0.5}
 \end{tikzpicture}$ along the indicated path~\cite{nahum2020entanglement,foligno2024nonequilibrium} and computing the value of the resulting diagram we find~\eqref{eq:semi_infinite_S}, thus confirming the entanglement membrane picture~\cite{Note11}.  This is also in agreement with the value of the operator entanglement along the same path.

\textit{Finite subsystem}---We now consider the case of a finite-size subsystem and restrict to the case where $r<\ell/2$, and also $\ell\ll L$ to avoid recurrences.  In this case the dynamics can be split into a number of different regimes, the simplest to analyze being $t\leq\ell/2$.  For that case the edges of the subsystem are not in causal contact with each other and $S^{(n)}_A(t)$ can be determined by summing up the contributions from each edge separately.  These  are obtained from~\eqref{eq:semi_infinite_S} with $S^{(n)}_{\rm imp}\to \log (q_l^2)$ for the edge farther
from the impurity.   

For $t>\ell/2$ the effect of the finite subsystem size can be felt.  Using the dual unitarity of the bulk we arrive at the following representation: 
\begin{align}
\label{eq:finite_size}
\!\!\!\tr[\rho^n_A(t)]=\hspace{-1.5cm} \begin{tikzpicture}[baseline={([yshift=0.6ex]current bounding box.center)},scale=0.45]
{\prop{2}{2}{colSt}{}}
   \foreach \x in {1,3}{\prop{\x}{1}{colSt}{}}   
    \foreach \x in {0,2,...,4}{\prop{\x}{0}{colSt}{}}
    \foreach \x in {-1,1,...,5}{\prop{\x}{-1}{colSt}{}}
    \foreach \x in {-2,0,...,6}{\prop{\x}{-2}{colSt}{}}
    \foreach \x in {-3,-1,...,5}{\prop{\x}{-3}{colSt}{}}
        \foreach \x in {7}{\newprop{\x}{-3}{colSt}{}}
    \foreach \x in {-4,-2,...,6}{\prop{\x}{-4}{colSt}{}}
     \foreach \x in {8}{\newprop{\x}{-4}{colSt}{}}
   \foreach \x in {-4,-2} {\propimp{6}{\x}{colimp}{}}
    \foreach \x in {-5.5,-3.5,...,4.5}{\istate{\x}{-4.5}{colSt}}
        \foreach \x in {6.5,8.5}{\newistate{\x}{-4.5}{colSt}}
    \foreach \x in {-4,...,1}{\square{\x+0.5}{\x+1.5}}
     \foreach \x in {-6,-5}{\circle{\x+0.5}{\x+1.5}}
    \foreach \x in {-6,...,1}{\circle
    {4-\x-0.5}{\x+1.5}}
      \draw[semithick,decorate,decoration={brace}] (-4,-2.25) -- (1.25,3) node[midway,xshift=-7.5pt,yshift=7.5pt] {$\ell$};
      \draw[semithick,decorate,decoration={brace}] (-6.2,-4.45) -- (-4.65,-2.9) node[midway,xshift=-11.5pt,yshift=7.5pt] {$2t-\ell$};
 \draw[semithick](9.5,-.25) -- (9.5,-.25) node[xshift=-21pt]{$\times \,q_l^{(1-n)\ell}$};
  \end{tikzpicture}
  \hspace{-.5cm}. 
\end{align}
When $t\geq \ell+r$, this diagram can be further reduced using both the solvability of the initial state~\eqref{eq:initial_state} and the unitarity of the local gates~\eqref{eq:local_gate_rels_1}. The result is that
\begin{eqnarray}\label{eq:saturation_value}
S^{(n)}_A(t\geq \ell+r)=\ell \log(q_l^2),
\end{eqnarray}
meaning that the subsystem relaxes to the infinite temperature state also in the presence of the impurity, although it is only guaranteed to do so at a much later time than that of a pure system, $t=\ell+r$ rather than $t=\ell/2$.  To investigate the intermediate times $\ell/2<t<\ell+r$, however, it is necessary to examine some specific examples. 

We start by considering the case where the circuit generates integrable dynamics, and in particular choose the simplest example where $U(x\neq L,t)=(u_+\otimes u_-) P(v_+\otimes v_-) $ with $u_\pm,v_\pm$ being single-site unitaries ($P$ is the swap gate).  With this choice we can straightforwardly reduce~\eqref{eq:finite_size} and find for $\ell/2<t<\ell+r$ that
\begin{equation}\label{eq:ent_swap_gates}
S^{(n)}_{A}\!(t)\!\!=\!\!\begin{cases}
(\ell+r-t)\log(q_l^2)+(t-r)S^{(n)}_{\rm imp} & t\le\frac{\ell}{2}+r,\\
(t-r)\log(q_l^2)+(\ell+r-t)S^{(n)}_{\rm imp}&\frac{\ell}{2}+r<t.
\end{cases}
\end{equation}
A notable feature of the above expression is that the entanglement decreases  between $\ell/2<t<\ell/2+r$  and then increases again before saturating at $t=\ell+r$, see Fig.~\ref{fig:ent}. We expect these features to be generic for integrable models as they admit an explanation via the quasiparticle picture~\cite{Note11}. In particular, the decrease in entanglement can be understood from the fact that the impurity can cause a quasiparticle to be reflected back inside the subsystem, thereby decreasing its overall contribution to the entanglement. After some time, however, the other quasiparticle of the pair  
will exit the subsystem from the other side and their contribution will be the same as in the pure case.

Thus $S^{(n)}_A(t)$ can exhibit non-monotonic behavior and a delayed saturation time for an integrable bulk circuit in the presence of an impurity. To investigate if these features can appear also in a non-integrable case, we introduce randomness to the bulk in the manner of~\cite{piroli2020scrambling} and also restrict to $n=2$ and $q_l=q_r= q$.  Upon averaging over different realizations, the bulk gates project onto the space spanned by the orthogonal states $\begin{tikzpicture}[baseline={([yshift=-0.6ex]current bounding box.center)},scale=0.5]
    \circle{-0.5}{0.5}
      \gridLine{-0.38}{0.5}{0.2}{0.5}
 \end{tikzpicture}$ and  $\begin{tikzpicture}[baseline={([yshift=-0.6ex]current bounding box.center)},scale=0.5]
    \circleblack{-0.5}{0.5}
      \gridLine{-0.38}{0.5}{0.2}{0.5}
 \end{tikzpicture}=(q\,\begin{tikzpicture}[baseline={([yshift=-0.6ex]current bounding box.center)},scale=0.5]
    \square{-0.5}{0.5}
      \gridLine{-0.38}{0.5}{0.2}{0.5}
 \end{tikzpicture}-\begin{tikzpicture}[baseline={([yshift=-0.6ex]current bounding box.center)},scale=0.5]
    \circle{-0.5}{0.5}
      \gridLine{-0.38}{0.5}{0.2}{0.5}
 \end{tikzpicture})/\sqrt{q^2-1}$. Within this space the averaged gate takes a simple form~\cite{foligno2023growth,Note11}, depending only on $q$ and on a parameter $p$, known as the entangling power of the gate~\cite{rather2020creating,jonnadula2020entangling,foligno2023growth}. This quantity, which we assume to be uniform among the bulk gates, obeys $0\leq p\leq 1$ and determines the mixing properties of the bulk, i.e., the decay rate of dynamical correlations~\cite{arvinda2021bernoulli}. For $p=0$ the averaged gate is the integrable swap gate, and light-cone correlations persist, while for $p=1$ the decay rate is infinite.

We start by analyzing the case of $p=1$ in the large $q$ limit. There it can be shown that~\cite{Note11}
\begin{eqnarray}\label{eq:maximal_ent}
S^{(2)}_A(t)=\begin{cases} \left(t+r\right)\log(q^2) +  (t-r)S^{(2)}_{\text{imp}}&t\leq t_*,\\
 \ell\log(q^2)&t>t_*,
\end{cases}
\end{eqnarray}
where $t_*\leq \ell-r$ is the saturation time,  defined so that $S^{(2)}_A(t)$ is continuous. The upper bound for $t_*$ occurs when the impurity is purely reflecting. 
Eq.~\eqref{eq:maximal_ent} can be reproduced using the entanglement membrane picture. To see this, we note that in the early time regime $t\leq \ell/2$ the entanglement membrane follows the red dashed lines indicated in Fig.~\ref{fig:rho_circuit}. On the other hand, for $t\geq\ell+r$ the exact result~\eqref{eq:saturation_value} suggests that the membrane instead runs horizontally across the subsystem, connecting the left and right edges. The expectation for the intermediate regime then would be that the membrane would continue to follow the path indicated in Fig.~\ref{fig:rho_circuit} until such time that the horizontal configuration is more favorable, in agreement with~\eqref{eq:maximal_ent} \cite{Note11}. 

When the bulk does not have maximal entangling power, i.e. $0<p<1$, the picture changes. Once again we work in the large $q$ limit such that $q\gg \left(1-p\right)^{-1}$. 
In the regime  $\ell/2<t\le\ell/2+r$ and  after averaging, one can then show that the purity obeys the following lower bound~\cite{Note11},
\be \label{eq:bound}
\tr[\rho^2_A(t)]\geq \frac{(1-p)^{\ell(2t-\ell)}}{q^{2\ell+2r-2t}}e^{(r-t)S^{(2)}_{\rm imp}}.
\ee
At $t=\ell/2$ the lower bound agrees with the exact calculation, whereas for $\log[ q^2(1-p)^{2\ell}]> S^{(2)}_{\rm imp}$, the right-hand side increases with $t$, indicating that the entanglement entropy \textit{must} decrease. 
At $p=0$ the lower bound gives the exact result of~\eqref{eq:ent_swap_gates}, while for a dual unitary impurity the inequality is satisfied by the saturation value~\eqref{eq:saturation_value}.  Moreover, in the regime  $\ell/2+r<t<\ell+r$ we find 
\be\label{eq:second_bound}
 \tr[\rho^2_A(t)]\!\geq\! \frac{1}{q^{2\ell}}+\frac{(1\!-\!p)^{2(\ell+r-t)(2t-\ell)}}{q^{2\ell}}\!\left(q^2 e^{-S_{\rm imp}^{\left(2\right)}}-1\right)^{\ell+r-t}\!\!\!\!\!\!, 
 \ee
 from which we can deduce that for $t<\ell+r$, 
 $S_A^{(2)}(t)$ is still growing, at least until the second term in~\eqref{eq:second_bound} becomes subleading in $1/q$ and can be omitted. 

Note that, at least to a leading order, the membrane picture is at odds with the possibility of decreasing entanglement due to the basic properties of the entanglement membrane tension~\cite{jonay2018coarse}. Thus, even though circuits with $0<p<1$ are generally chaotic (since integrable models only constitute a zero-measure subset~\cite{bertini2020operator}), their entanglement dynamics may disagree with the membrane picture for a finite subsystem. Given any $p>0$, for a large enough $\ell$ the bound~\eqref{eq:bound} ceases to ensure the entropy decrease, indicating that the membrane picture could indeed emerge above a certain crossover scale. Nevertheless, this bound reveals the scale below which a more fine-grained theory is required.

\textit{Conclusions}---
In this work we have examined the growth of bipartite entanglement in the presence of impurities in otherwise dual unitary quantum circuits.  We have shown that for semi-infinite subsystems the exact result can be interpreted using either the quasiparticle or membrane pictures, while for finite-size subsystems the two theories give differing predictions. We found that (early-time) entanglement dynamics in a swap-bulk circuit can be exactly captured by a quasiparticle picture, despite the impurity breaking integrability. We also revealed a class of chaotic systems that, below a certain crossover scale, are not described by the membrane picture, raising the need for formulating a fine-grained theory interpolating between it and the quasiparticle picture (see ~\cite{jonay2023physical} for recent work along these lines). Such an effort could be assisted by a systematic numerical study of the averaged random circuits we discussed.  Another interesting continuation of this work would be to consider generalizations of these circuits to other gates in the dual unitary hierarchy~\cite{yu2024hierarchy}, charged dual unitary circuits~\cite{foligno2024nonequilibrium}, or Haar-random circuits~\cite{PhysRevB.98.035118}.

\textit{Acknowledgments}--- The authors wish to thank B.~Bertini, P.~Calabrese, M.~Goldstein, K.~Klobas, and T.~Zhou for useful discussions. C.R.~acknowledges support from ERC under Consolidator Grant number 771536 (NEMO) and the European Union - NextGenerationEU, in the framework of the PRIN Project HIGHEST number 2022SJCKAH\_002. S.F.~is grateful for the support of the Azrieli Foundation Fellows program and of the Tel-Aviv University Center for Nanoscience and Nanotechnology.

\bibliography{Dual_Unitary_Imp}







\onecolumngrid
\newpage 
\newcounter{equationSM}
\newcounter{figureSM}
\newcounter{tableSM}
\stepcounter{equationSM}
\setcounter{equation}{0}
\setcounter{figure}{0}
\setcounter{table}{0}
\setcounter{section}{0}
\makeatletter
\renewcommand{\theequation}{\textsc{sm}-\arabic{equation}}
\renewcommand{\thefigure}{\textsc{sm}-\arabic{figure}}
\renewcommand{\thetable}{\textsc{sm}-\arabic{table}}

\begin{center}
  {\large{\bf Supplemental Material for ``Entanglement in dual unitary quantum circuits with impurities''}}
\end{center}



Here we report some useful information complementing the main text. In particular:
\begin{itemize}
   \item[-] In Sec.~\ref{sec:correlation} we discuss the correlation functions in the presence of the impurity and the transmission and reflection coefficients.
   \item[-] In Sec.~\ref{sec:Quasiparticle_picture} we provide a quasiparticle-picture explanation for the entanglement dynamics of integrable circuits perturbed by an impurity.
   \item[-] In Sec.~\ref{sec:membrane_picture} we present details of the prediction for the entanglement dynamics using the membrane picture approach.
   \item[-] In Sec.~\ref{sec:maximal} we present details of the exact entropy calculation for random gates with maximal entangling power.
   \item[-] In Sec.~\ref{sec:bounds} we derive the bounds on the purity presented in the main text.
\end{itemize}

\section{Correlations}\label{sec:correlation}
 For pure dual unitary brickwork quantum circuits, it has been shown that correlation functions are non-trivial only between points connected by light rays~\cite{bertini2019exact}. When some of the gates are perturbed from dual unitarity, however, this special structure is lost and correlation functions involve sums over all allowed paths through the system~\cite{kos2021correlations}. For the particular system we study, the situation is much simpler. Infinite temperature correlations can occur only between points connected along light rays or those which are connected by a ray that reflects off the impurity,  see Fig.~\ref{fig:correlation_funcs}.  

It is instructive to see this in action for a specific choice of an impurity gate. We take $q_l=q_r=q$ and choose
\be\label{eq:Kondo_imp}
U(L,t)=\frac{\1+i c P}{1+i c},
\ee
where $c\in\mathbb{R}$ and $P$ is the swap gate on two qudits, which acts as 
$P\ket{ij}=\ket{ji} $. The form of this impurity gate is quite natural as it tunes between a completely reflecting impurity at $c=0$, which does not allow for the transmission of information, and a completely transparent one at $c\to\infty$, which is dual unitary.  

Let us consider the infinite temperature 2-point correlation function $\mathcal{C}^{\alpha,\beta}(x,y;t)\equiv\tr[\sigma^{\alpha}(x,t)\sigma^\beta(y,0)]$, where $\sigma^{\alpha,\beta}(z,\tau)$ are traceless single-site operators at space-time positions $z,\tau$.  Setting $y\geq L+1$ with $y-L$ being odd, the computation of this quantity can be reduced to the diagrams in Fig.~\ref{fig:correlation_funcs} (a) and (b), where it is understood that there is only a single replica, $n=1$.  Using \eqref{eq:Kondo_imp} we find that 
\begin{eqnarray}
 \mathcal{C}^{\alpha,\beta}(y-2t,y;t)=\mathcal{T} \lim_{c\to \infty} \mathcal{C}^{\alpha,\beta}(y-2t,y;t),\\
   \mathcal{C}^{\alpha,\beta}(2t-y+2L+1,y;t)=\mathcal{R}\lim_{c\to \infty} \mathcal{C}^{\alpha,\beta}(y-2t,y;t),
\end{eqnarray}
where $\mathcal{T}=\frac{c^2}{1+c^2}$, $\mathcal{R}=1-\mathcal{T}$ are the transmission and reflection coefficients
for the impurity. These multiply a factor that corresponds to the correlation function in the presence of a completely transparent impurity.  Thus, irrespective of the nature of the bulk, information can be transmitted or reflected from the impurity. 
The same coefficients then appear in the entanglement entropy in Eq.~(16) of the main text.
For the case of a quench, the correlation functions can no longer be reduced to this simple form due to the impurity not being dual unitary, see Fig.~\ref{fig:correlation_funcs} (c) and (d). As a result, they depend on higher powers of $\mathcal{T}$ and $\mathcal{R}$, indicating that multiple scattering processes are involved. 

\begin{figure}[H]\center
(a) \,\,\begin{tikzpicture}[baseline={([yshift=10ex]current bounding box.center)},scale=0.45]
     \foreach \x in {4,...,6}{\prop{\x}{6-\x}{colSt}{}}
       \foreach \x in {8,...,11}{\prop{\x}{6-\x}{colSt}{}}
     {\prop{7}{-1}{colimp}{}}
      \foreach \x in {4,...,11}{\circle{\x+.5}{6.5-\x}}
       \foreach \x in {4,...,11}{\circle{\x-.5}{5.5-\x}}
       {\circler{11.5}{-5.5}}
       {\circleg{3.5}{2.5}}
  \draw[semithick,decorate,decoration={brace}] (4.5,3) -- (12,-4.5) node[midway,xshift=12.5pt] {$2t$};
  \end{tikzpicture}
 \,\,(b) \,\,\begin{tikzpicture}[baseline={([yshift=10ex]current bounding box.center)},scale=0.45]
     \foreach \x in {8,...,11}{\prop{\x}{6-\x}{colSt}{}}
       \foreach \x in {8,...,10}{\prop{\x}{\x-8}{colSt}{}}
     {\prop{7}{-1}{colimp}{}}
     \foreach \x in {8,...,11}{\circle{\x+.5}{6.5-\x}}
       \foreach \x in {7,...,11}{\circle{\x-.5}{5.5-\x}}
        \foreach \x in {7,...,10}{\circle{\x-.5}{-7.5+\x}}
        \foreach \x in {8,...,10}{\circle{\x+.5}{-8.5+\x}}
          {\circler{11.5}{-5.5}}
            {\circleg{10.5}{2.5}}
  \end{tikzpicture}
  (c) \begin{tikzpicture}[baseline={([yshift=10ex]current bounding box.center)},scale=0.45]
     \foreach \x in {4,...,6}{\prop{\x}{6-\x}{colSt}{}}
       \foreach \x in {8,...,11}{\prop{\x}{6-\x}{colSt}{}}  
        \foreach \x in {0,-2,-4} {\prop{7}{\x-1}{colimp}{}}
      \foreach \x in {4,...,11}{\circle{\x+.5}{6.5-\x}}
       \foreach \x in {4,...,7}{\circle{\x-.5}{5.5-\x}}
        \foreach \x in {1,...,4}{\circle{6.5}{-1.5-\x}}
       {\circleg{3.5}{2.5}}
       \foreach \x in {7.5,9.5,11.5} {\istate{\x}{-5.5}{colSt}}
       \prop{9}{-5}{colSt}{}
        \prop{8}{-4}{colSt}{}
          {\circler{12.5}{-5.5}}
  \end{tikzpicture}
 \,\,(d) \,\,\begin{tikzpicture}[baseline={([yshift=10ex]current bounding box.center)},scale=0.45]
     \foreach \x in {8,...,11}{\prop{\x}{6-\x}{colSt}{}}
       \foreach \x in {8,...,10}{\prop{\x}{\x-8}{colSt}{}}
   \foreach \x in {0,-2,-4} {\prop{7}{\x-1}{colimp}{}}
     \foreach \x in {8,...,11}{\circle{\x+.5}{6.5-\x}}
       \foreach \x in {7}{\circle{\x-.5}{5.5-\x}}
        \foreach \x in {7,...,10}{\circle{\x-.5}{-7.5+\x}}
        \foreach \x in {8,...,10}{\circle{\x+.5}{-8.5+\x}}
         \foreach \x in {7.5,9.5,11.5} {\istate{\x}{-5.5}{colSt}}
            \prop{9}{-5}{colSt}{}
        \prop{8}{-4}{colSt}{}
          \foreach \x in {1,...,4}{\circle{6.5}{-1.5-\x}}
            {\circleg{10.5}{2.5}}
            {\circler{12.5}{-5.5}}
  \end{tikzpicture}
  \caption{\label{fig:correlation_funcs} (a,b) Diagrammatic representation of infinite temperature correlation functions $\tr[\sigma^\alpha (x,t)\sigma^\beta(y,0)]$ in the presence of an impurity.  Operators  $\sigma^{\beta}$, $\sigma^{\alpha}$ are denoted by red and green circles, respectively, and the local gates correspond to the single replica case $n=1$.   Correlation functions are supported along (a) light rays between points or (b) rays that reflect off the impurity. Depicted is the case of $t=4$ , $y=L+5$, and (a) $x=L-3$ or (b) $x=L+4$. (c,d) Correlation functions after a quench for the same parameters. }
\end{figure}

\section{Quasiparticle picture of entanglement dynamics}\label{sec:Quasiparticle_picture}
In this section, we explain how the results in Eqs.~(14) and (20) of the main text, which state
the entanglement entropy evolution in an integrable circuit perturbed by an impurity, are captured by a coarse-grained quasiparticle picture description. In particular, we examine entanglement dynamics ensuing from pairs of maximally-entangled counter-propagating quasiparticles generated uniformly across the system at $t=0$. To match the circuit defined in the main text, their density is such that a single pair is generated in any interval of length $\Delta x=2$, and the quasiparticle velocity is $v=2$. The quasiparticles propagate independently almost everywhere, and can interact only at the location of the impurity, $x=L$.

Therefore, to obtain a quasiparticle picture description, one must determine the effect of the impurity on the entanglement between quasiparticles. Consider a quasiparticle pair generated at a point $x=x_0<L$ to the left of the impurity, and let $A^{(+)}_{x_0}$ and $A^{(-)}_{x_0}$ denote the single-quasiparticle subsystems of the right-moving and left-moving parts of the pair, respectively. At time $t=\left(L-x_0\right)/v$, the right-moving quasiparticle reaches the impurity, simultaneously with the left-moving quasiparticle of the pair generated at $x=2L-x_0$, and they are both reflected off the impurity. The two quasiparticle pairs become entangled due to the impurity, the action of which corresponds to the unitary $U(L,t)$. By the definition of operator entanglement~\cite{zanardi2001entanglement,rampp2023spreading}, we then have that the entropy of the combined subsystem $A^{(+)}_{x_0}\cup A^{(-)}_{x_0}$ is
\be
S^{(n)}_{A^{(+)}_{x_0}\cup A^{(-)}_{x_0}}=S^{(n)}_{\rm imp},
\ee
where $S^{(n)}_{\rm imp}$ was defined in Eq.~(15) of the main text. 
That is, the two \emph{pairs} on the opposite sides of the impurity are entangled to each other to an extent determined by $S^{(n)}_{\rm imp}$.

\begin{figure}
  \includegraphics[trim={1cm 4cm 1cm 0}, clip,width=1\textwidth]{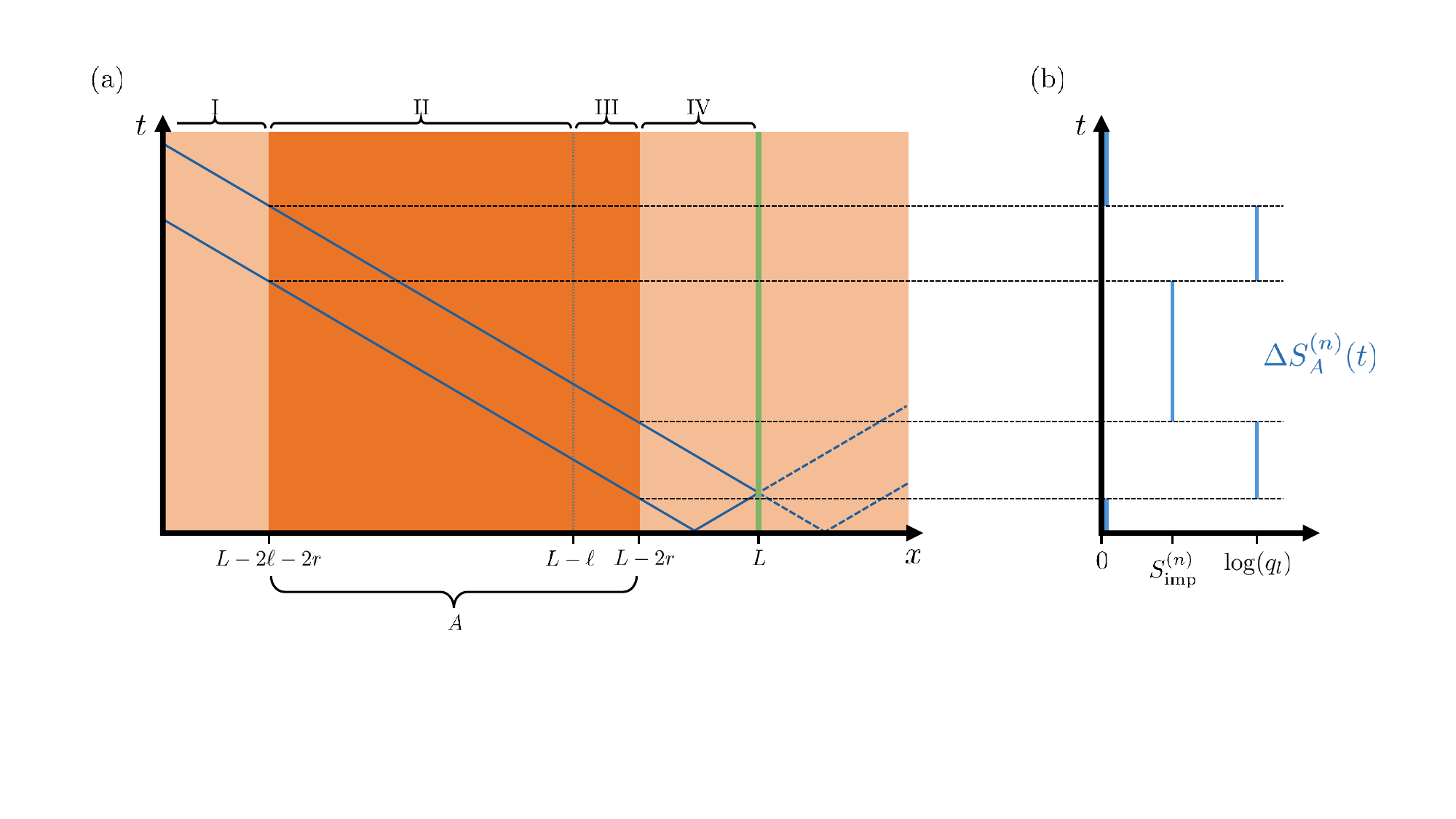}
  \caption{\label{fig:QP_picture_SM} The quasiparticle picture of entanglement dynamics in the presence of an impurity. (a) Coarse-grained space-time plot of the system, with an example for a pair of correlated counter-propagating particles generated at $t=0$ on the left side of the impurity (solid blue lines); at a certain point in time, the right-moving quasiparticle reaches the impurity at $x=L$ (solid vertical line), and becomes entangled with a quasiparticle from a pair generated on the right side of the impurity (dashed blue lines); then both quasiparticles are reflected off the impurity. The plot also shows the location of the finite subsystem $A$ defined in the main text, and the division of the domain to the left of the impurity into the four regions discussed in Sec.~\ref{sec:Quasiparticle_picture}. (b) The contribution to the entanglement entropy of $A$ arising from the exemplary quasiparticle pair as a function of time.}
\end{figure}

Therefore, if a subsystem $A$ contains at some point $t$ in time the two quasiparticles generated at $x=x_0$, their contribution to its entropy $S^{(n)}_A$ will be $\Delta S^{(n)}_A=0$ for $t<\left(L-x_0\right)/v$ and $\Delta S^{(n)}_A=S^{(n)}_{\rm imp}$ for $t>\left(L-x_0\right)/v$. If $A$ contains only one of these quasiparticles, at any point in time its contribution to $S^{(n)}_A$ will be maximal, i.e., $\Delta S^{(n)}_A=\log(q_l)$. This is clear if the quasiparticle in question is the left one (which never interacts with another quasiparticle), or the right one for $t<\left(L-x_0\right)/v$. The fact that it is true also for the right quasiparticle when $t>\left(L-x_0\right)/v$ is observed by noting that the two quasiparticles that meet at the impurity constitute a subsystem with maximal entropy; since the impurity operator $U(L,t)$ cannot reduce this entropy, they maintain their maximal entropy, meaning that each one of them must be maximally entangled to the rest of the system. A schematic example for the entropy contribution of a single quasiparticle pair is depicted in Fig.~\ref{fig:QP_picture_SM}.

What remains now is to add up the contributions of pairs generated at $t=0$ across the system. Our computation for a single pair implies that we should take into account only pairs generated at positions $x<L$, since their independent contributions already take into account the interactions with pairs generated at positions $x>L$. We divide the domain $x<L$ into four regions, as shown in Fig.~\ref{fig:QP_picture_SM}. For pairs generated in Region I, that is, the region $(0,L-2\ell-2r)$ outside $A$, only their right-moving part can be found inside $A$, with maximal entanglement contribution. The entanglement entropy arising from quasiparticles generated in this region is therefore
\begin{equation}\label{eq:EntropyRegion1}
S^{(n,{\rm I})}_{A}(t)=\begin{cases}
t\log(q_l) & 0\le t<\ell,\\
\ell\log(q_l)&\ell\le t < \ell+2r,\\
(t-2r)\log(q_l)&\ell+2r\le t < 2\ell+2r,\\
2\ell\log(q_l)&2\ell+2r\le t.
\end{cases}
\end{equation}
For pairs generated in Region II, defined as $(L-2\ell-2r,L-\ell)$, it is easy to check that the two quasiparticles of the same pair can be inside $A$ simultaneously only before the right-moving one exits $A$ for the first time, so they have a nonzero contribution to $S^{(n)}_A$ only when $A$ contains just one of them. This yields the contribution
\begin{equation}
S^{(n,{\rm II})}_{A}(t)=\begin{cases}
t\log(q_l) & 0\le t<\frac{\ell}{2}-r,\\
(2t+r-\frac{\ell}{2})\log(q_l)&\frac{\ell}{2}-r\le t<\frac{\ell}{2},\\
(\frac{3}{2}\ell+r-2t)\log(q_l)&\frac{\ell}{2}\le t < \frac{\ell}{2}+r,\\
(\frac{\ell}{2}-r)\log(q_l)&\frac{\ell}{2}+r\le t < \ell,\\
(t-\frac{\ell}{2}-r)\log(q_l)&\ell\le t < \ell+2r,\\
(\frac{\ell}{2}+r)\log(q_l)& \ell+2r\le t < \frac{3}{2}\ell+r,\\
(2\ell+2r-t)\log(q_l)&\frac{3}{2}\ell+r\le t <2\ell+2r,\\
0&2\ell+2r\le t.
\end{cases}
\end{equation}

\begin{figure}
  \includegraphics[
  width=0.7\textwidth]
  {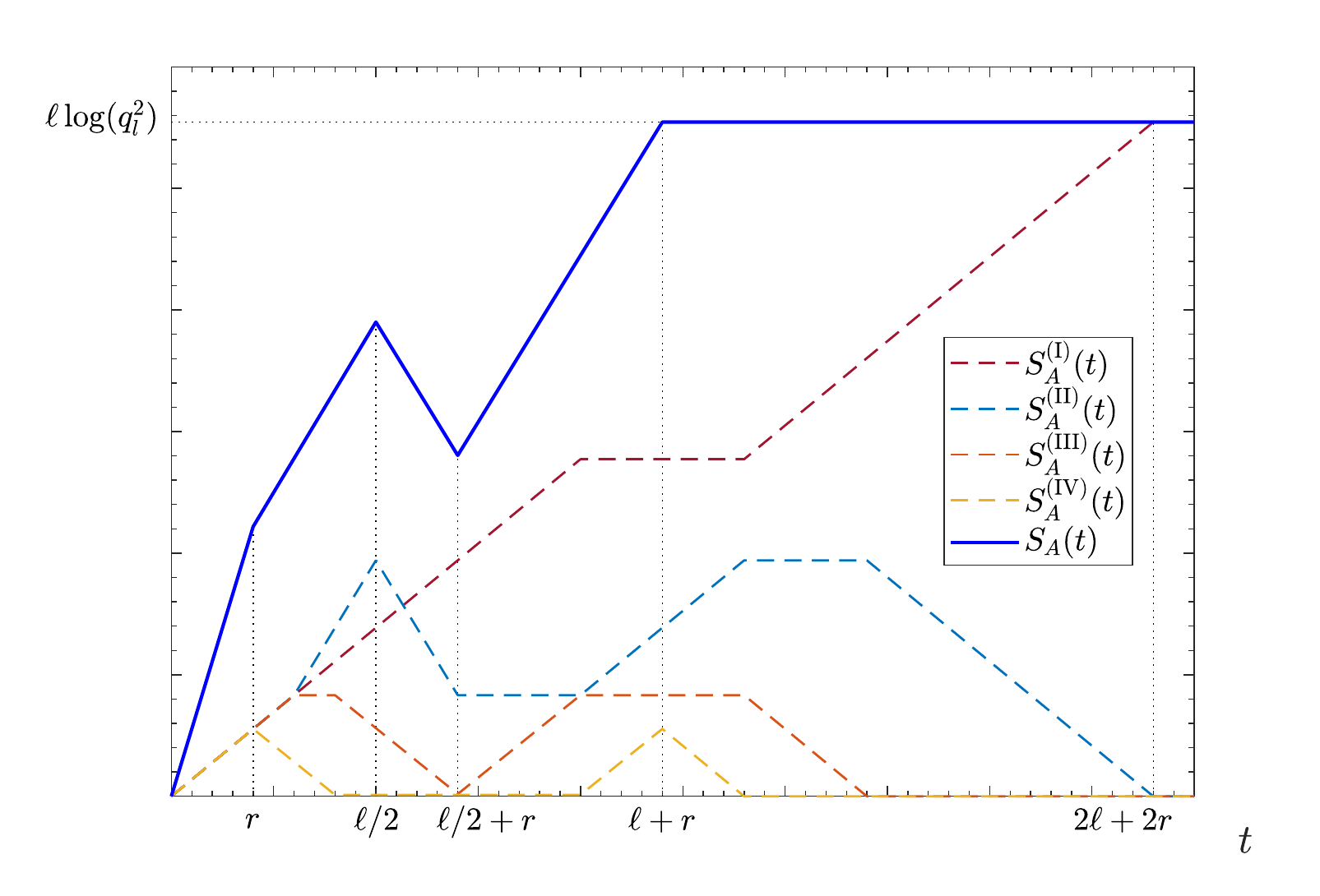}
  \caption{\label{fig:QP_picture_contributions_SM} Entanglement entropy dynamics in the bulk swap-gate circuit perturbed by an impurity. Subsystem $A$ has length $2\ell$ and is a distance $2r$ from the impurity. Results are shown for local Hilbert space dimension $q_l=2$ and the impurity given in Eq.~(16) of the main text, with $c=0.05$, in the replica limit $n\to 1$. The plot shows the contributions arising from Regions I-IV (Eqs.~\eqref{eq:EntropyRegion1}--\eqref{eq:EntropyRegion4}) as well as the total entropy (Eq.~\eqref{eq:TotalQPEntropy}).}
\end{figure}

In contrast to the cases of Regions I and II, quasiparticles of the same pair generated in $(L-\ell,L)$ can, during a certain time span, be simultaneously inside $A$ even after the right quasiparticle is reflected off the impurity, contributing $S^{(n)}_{\rm imp}$ to $S^{(n)}_A$ during that time. Adding up contributions from quasiparticles generated in Region III, namely the region $(L-\ell,L-2r)$ contained in $A$, and in Region IV, namely the region $(L-2r,L)$ between $A$ and the impurity, one obtains the following:
\begin{equation}
S^{(n,{\rm III})}_{A}(t)=\begin{cases}
t\log(q_l) & 0\le t<\min\{\frac{\ell}{2}-r,2r\},\\
\min\{\frac{\ell}{2}-r,2r\}\log(q_l)+\max\{t-2r,0\}S^{(n)}_{\rm imp}&\min\{\frac{\ell}{2}-r,2r\}\le t<\max\{\frac{\ell}{2}-r,2r\},\\
(\frac{\ell}{2}+r-t)\log(q_l)+(t-2r)S^{(n)}_{\rm imp}&\max\{\frac{\ell}{2}-r,2r\}\le t < \frac{\ell}{2}+r,\\
(t-\frac{\ell}{2}-r)\log(q_l)+(\ell-t)S^{(n)}_{\rm imp}&\frac{\ell}{2}+r\le t <\ell,\\
(\frac{\ell}{2}-r)\log(q_l)&\ell\le t < \ell+2r,\\
(\frac{3}{2}\ell+r-t)\log(q_l)&\ell+2r\le t <\frac{3}{2}\ell+r,\\
0&\frac{3}{2}\ell+r\le t,
\end{cases}
\end{equation}
and
\begin{equation}\label{eq:EntropyRegion4}
S^{(n,{\rm IV})}_{A}(t)=\begin{cases}
t\log(q_l) & 0\le t<r,\\
(2r-t)\log(q_l)+(t-r)S^{(n)}_{\rm imp} & r\le t <2r,\\
rS^{(n)}_{\rm imp}&2r\le t <\ell,\\
(t-\ell)\log(q_l)+(\ell+r-t)S^{(n)}_{\rm imp} &\ell\le t< \ell+r,\\
(\ell+2r-t)\log(q_l) & \ell+r\le t <\ell+2r,\\
0&\ell+2r\le t.
\end{cases}
\end{equation}

Finally, the total entanglement entropy of $A$ is obtained from summing up the contributions arising from the different regions. This leads to
\begin{equation}\label{eq:TotalQPEntropy}
S^{(n)}_{A}(t)=S^{(n,{\rm I})}_{A}(t)+S^{(n,{\rm II})}_{A}(t)+S^{(n,{\rm III})}_{A}(t)+S^{(n,{\rm IV})}_{A}(t)=\begin{cases}
2t\log(q_l^2) & 0\le t<r,\\
(t+r)\log(q_l^2)+(t-r)S^{(n)}_{\rm imp} & r\le t <\frac{\ell}{2},\\
(\ell+r-t)\log(q_l^2)+(t-r)S^{(n)}_{\rm imp}&\frac{\ell}{2}\le t <\frac{\ell}{2}+r,\\
(t-r)\log(q_l^2)+(\ell+r-t)S^{(n)}_{\rm imp} &\frac{\ell}{2}+r\le t< \ell+r,\\
\ell\log(q_l^2)&\ell+r\le t,
\end{cases}
\end{equation}
which indeed fits the result for the swap-gate circuit presented in the main text, see Eqs.~(14) and (20). 
Fig.~\ref{fig:QP_picture_contributions_SM} shows the same example that appeared in Fig.~1 of the main text of the entanglement entropy dynamics predicted by the quasiparticle picture, along with the different contributions arising from Regions I--IV.

\section{Entanglement membrane picture of entanglement dynamics}\label{sec:membrane_picture}
In this section, we provide details on the entanglement membrane approach to calculating the dynamics of $\tr[\rho_A(t)^n]$ in our model. In diagrams this can be depicted as 

\begin{align}
\!\!\!\tr[\rho^n_A(t)]= \begin{tikzpicture}[baseline={([yshift=0.6ex]current bounding box.center)},scale=0.45]
{\prop{2}{2}{colSt}{}}
   \foreach \x in {1,3}{\prop{\x}{1}{colSt}{}}   
    \foreach \x in {0,2,...,4}{\prop{\x}{0}{colSt}{}}
    \foreach \x in {-1,1,...,5}{\prop{\x}{-1}{colSt}{}}
    \foreach \x in {-2,0,...,6}{\prop{\x}{-2}{colSt}{}}
    \foreach \x in {-3,-1,...,5}{\prop{\x}{-3}{colSt}{}}
        \foreach \x in {7}{\newprop{\x}{-3}{colSt}{}}
    \foreach \x in {-4,-2,...,6}{\prop{\x}{-4}{colSt}{}}
     \foreach \x in {8}{\newprop{\x}{-4}{colSt}{}}
   \foreach \x in {-4,-2} {\propimp{6}{\x}{colimp}{}}
    \foreach \x in {-5.5,-3.5,...,4.5}{\istate{\x}{-4.5}{colSt}}
        \foreach \x in {6.5,8.5}{\newistate{\x}{-4.5}{colSt}}
    \foreach \x in {-4,...,1}{\square{\x+0.5}{\x+1.5}}
     \foreach \x in {-6,-5}{\circle{\x+0.5}{\x+1.5}}
    \foreach \x in {-6,...,1}{\circle
    {4-\x-0.5}{\x+1.5}}
      \draw[semithick,decorate,decoration={brace}] (-4,-2.25) -- (1.25,3) node[midway,xshift=-7.5pt,yshift=7.5pt] {$\ell$};
      \draw[semithick,decorate,decoration={brace}] (-6.2,-4.45) -- (-4.65,-2.9) node[midway,xshift=-11.5pt,yshift=7.5pt] {$2t-\ell$};
 \draw[semithick](9.5,-.25) -- (9.5,-.25) node[xshift=-1pt]{$\times \,q_l^{(1-n)\ell}$};
\draw[semithick,decorate,decoration={brace}] (2.7,3.3) -- (6,-0.05) node[midway,xshift=7.5pt,yshift=7.5pt] {$2r$};
  \end{tikzpicture}
  \hspace{-.5cm}. 
\end{align}
Here we depict the case for $t>\ell/2$ but the case $t<\ell/2$ follows from a similar calculation.  The main idea behind the approach is that the leading contribution to the above diagram comes from projecting each leg of this diagram onto either circle or square states, which form domains. Interfaces between these domains are the eponymous membranes, and their free energy gives the entanglement entropy. 

Practically, the calculation proceeds by picking a downward-directed path for the membrane originating at the top of the diagram where there is an interface between the circle and square states, and either ending on the initial state or joining another membrane, see Fig.~1 of the main text. For the above diagram, we choose the membrane to follow the red dashed lines depicted below,
\begin{align}
    \begin{tikzpicture}[baseline={([yshift=0.6ex]current bounding box.center)},scale=0.45]
{\prop{2}{2}{colSt}{}}
   \foreach \x in {1,3}{\prop{\x}{1}{colSt}{}}   
    \foreach \x in {0,2,...,4}{\prop{\x}{0}{colSt}{}}
    \foreach \x in {-1,1,...,5}{\prop{\x}{-1}{colSt}{}}
    \foreach \x in {-2,0,...,6}{\prop{\x}{-2}{colSt}{}}
    \foreach \x in {-3,-1,...,5}{\prop{\x}{-3}{colSt}{}}
        \foreach \x in {7}{\newprop{\x}{-3}{colSt}{}}
    \foreach \x in {-4,-2,...,6}{\prop{\x}{-4}{colSt}{}}
     \foreach \x in {8}{\newprop{\x}{-4}{colSt}{}}
   \foreach \x in {-4,-2} {\propimp{6}{\x}{colimp}{}}
    \foreach \x in {-5.5,-3.5,...,4.5}{\istate{\x}{-4.5}{colSt}}
        \foreach \x in {6.5,8.5}{\newistate{\x}{-4.5}{colSt}}
    \foreach \x in {-4,...,1}{\square{\x+0.5}{\x+1.5}}
     \foreach \x in {-6,-5}{\circle{\x+0.5}{\x+1.5}}
    \foreach \x in {-6,...,-3}{\circle
    {4-\x-0.5}{\x+1.5}}
    \foreach \x in {-2,...,1}{
    \tgridLine{4-\x-0.25}{\x+1.75}{4-\x-0.5}{\x+1.5}\circle
    {4-\x-0.25}{\x+1.75}}
      \draw[semithick,decorate,decoration={brace}] (-4,-2.25) -- (1.25,3) node[midway,xshift=-7.5pt,yshift=7.5pt] {$\ell$};
      \draw[semithick,decorate,decoration={brace}] (-6.2,-4.45) -- (-4.65,-2.9) node[midway,xshift=-11.5pt,yshift=7.5pt] {$2t-\ell$};
\draw[semithick,decorate,decoration={brace}] (2.7,3.3) -- (6,-0.05) node[midway,xshift=7.5pt,yshift=7.5pt] {$2r$};
\draw[very thick,dashed,red] (2,3) -- (5.5,-0.5) -- (5.5,-5) ;
      \draw[very thick,dashed,red] (-3.75,-3.25) -- (-2,-5) ;
  \end{tikzpicture}
  \hspace{-.5cm}. 
\end{align}
The membrane picture posits that the region between the membranes is in the domain of squares, while the regions to the left or right are within the circle domain. 
In this instance the domain of squares is dominant, since the number of gates between the membranes outnumbers those outside of them.  The contribution of the membrane is found by inserting a projector onto the square states along any link which crosses this path. Specifically, we replace cut links as follows 
\begin{eqnarray}
\begin{tikzpicture}[baseline={([yshift=-0.6ex]current bounding box.center)},scale=0.65]
 \gridLine{-0.5}{0}{0.5}{0}\draw[very thick,dashed,red] (0,0.4) -- (0,-.45);
 \end{tikzpicture}\rightarrow 
\begin{tikzpicture}[baseline={([yshift=-0.6ex]current bounding box.center)},scale=0.65]
      \square{0.5}{-0.5}
        \gridLine{0.62}{-0.5}{1}{-0.5}
         \square{-0}{-0.5}
        \gridLine{-0.12}{-0.5}{-0.5}{-0.5}
 \end{tikzpicture}~,
\end{eqnarray}
 and then account for the new normalization that this introduces by dividing by $q^{n{L_m}}$ where $L_m$ is the total length of the membranes, which is equal to the number of projectors inserted.  Carrying out this procedure, we find that the diagram becomes
\begin{align}
    \begin{tikzpicture}[baseline={([yshift=-5ex]current bounding box.center)},scale=0.45]
{\prop{2}{2}{colSt}{}}
   \foreach \x in {1,3}{\prop{\x}{1}{colSt}{}}   
    \foreach \x in {0,2,...,4}{\prop{\x}{0}{colSt}{}}
    \foreach \x in {-1,1,...,5}{\prop{\x}{-1}{colSt}{}}
    \foreach \x in {-2,0,...,4}{\prop{\x}{-2}{colSt}{}}
    \foreach \x in {-3,-1,...,5}{\prop{\x}{-3}{colSt}{}}
    \foreach \x in {-2,0,...,4}{\prop{\x}{-4}{colSt}{}}
    \foreach \x in {-1.5,0.5,...,4.5}{\istate{\x}{-4.5}{colSt}}
  \foreach \x in {-7.5,-5.5}{\istate{\x}{-4.5}{colSt}}
  \prop{-6}{-4}{colSt}{}
  \circle{-7.5}{-4.5}
  \circle{-6.5}{-3.5}
  \square{-5.5}{-3.5}
  \square{-4.5}{-4.5}
    \foreach \x in {-4,...,1}{\square{\x+0.5}{\x+1.5}}
    \square{-3.5}{-3.5}
     \square{-2.5}{-4.5}
    \foreach \x in {-2,...,1}{
    \square
    {4-\x-0.5}{\x+1.5}}
       \foreach \x in {-2,...,1}{
    \tgridLine {4.5-\x-0.5}{\x+2} {5-\x-0.5}{\x+2.5} \circle
    {5-\x-0.5}{\x+2.5}
     \square
    {4.5-\x-0.5}{\x+2}}
    \foreach \x in {1,...,4}
    {\square{5.5}{-0.5-\x}}
     \foreach \x in {-4,-2} {\propimp{7}{\x}{colimp}{}}
     \newprop{9}{-4}{colSt}{}
      \newprop{8}{-3}{colSt}{}
     \newistate{7.5}{-4.5}{colSt}
      \newistate{9.5}{-4.5}{colSt}
          \foreach \x in {1,...,4}
    {\square{6.5}{-0.5-\x}}
         \foreach \x in {1,...,4}
    {\circle{6.5+\x}{-0.5-\x}}
      \draw[semithick,decorate,decoration={brace}] (-4,-2.25) -- (1.25,3) node[midway,xshift=-7.5pt,yshift=7.5pt] {$\ell$};
       \draw[semithick,decorate,decoration={brace}] (-8.2,-4.45) -- (-6.65,-2.9) node[midway,xshift=-11.5pt,yshift=7.5pt] {$2t-\ell$};\draw[semithick,decorate,decoration={brace}] (3.75,4.25) -- (7,1) node[midway,xshift=7.5pt,yshift=7.5pt] {$2r$};
  \end{tikzpicture}. 
\end{align}
This diagram can be fully contracted, and upon accounting for the normalization factors, we obtain the result in Eq.~(21) of the main text for $t\leq t_*$. 

For times $t>t_*$, we instead take the membranes to traverse the subsystem horizontally and join together  as depicted below,
\begin{align}
    \begin{tikzpicture}[baseline={([yshift=0.6ex]current bounding box.center)},scale=0.45]
{\prop{2}{2}{colSt}{}}
   \foreach \x in {1,3}{\prop{\x}{1}{colSt}{}}   
    \foreach \x in {0,2,...,4}{\prop{\x}{0}{colSt}{}}
    \foreach \x in {-1,1,...,5}{\prop{\x}{-1}{colSt}{}}
    \foreach \x in {-2,0,...,6}{\prop{\x}{-2}{colSt}{}}
    \foreach \x in {-3,-1,...,5}{\prop{\x}{-3}{colSt}{}}
        \foreach \x in {7}{\newprop{\x}{-3}{colSt}{}}
    \foreach \x in {-4,-2,...,6}{\prop{\x}{-4}{colSt}{}}
     \foreach \x in {8}{\newprop{\x}{-4}{colSt}{}}
   \foreach \x in {-4,-2} {\propimp{6}{\x}{colimp}{}}
    \foreach \x in {-5.5,-3.5,...,4.5}{\istate{\x}{-4.5}{colSt}}
        \foreach \x in {6.5,8.5}{\newistate{\x}{-4.5}{colSt}}
    \foreach \x in {-4,...,1}{\tgridLine{\x+.25}{\x+1.75}{\x+.5}{\x+1.5}
    \square{\x+.25}{\x+1.75}}
     \foreach \x in {-6,-5}{\circle{\x+0.5}{\x+1.5}}
    \foreach \x in {-6,...,1}{\circle
    {4-\x-0.5}{\x+1.5}}
      \draw[semithick,decorate,decoration={brace}] (-4,-2.15) -- (1.25,3.15) node[midway,xshift=-7.5pt,yshift=7.5pt] {$\ell$};
      \draw[semithick,decorate,decoration={brace}] (-6.2,-4.45) -- (-4.65,-2.9) node[midway,xshift=-11.5pt,yshift=7.5pt] {$2t-\ell$};
\draw[semithick,decorate,decoration={brace}] (2.7,3.3) -- (6,-0.05) node[midway,xshift=7.5pt,yshift=7.5pt] {$2r$};
\draw[very thick,dashed,red] (2,3)--(-3.75,-2.75) ;
  \end{tikzpicture}
  \hspace{-.5cm}. 
\end{align}
The dominant region in this diagram is now that of the circle states which occupy the space to the right and below the membrane. To compute the contribution, we proceed in a similar fashion, however in this case we should project the legs onto circle states instead, 
\begin{eqnarray}
\begin{tikzpicture}[baseline={([yshift=-0.6ex]current bounding box.center)},scale=0.65]
 \gridLine{-0.5}{0}{0.5}{0}\draw[very thick,dashed,red] (0,0.4) -- (0,-.45);
 \end{tikzpicture}\rightarrow 
\begin{tikzpicture}[baseline={([yshift=-0.6ex]current bounding box.center)},scale=0.65]
      \circle{0.5}{-0.5}
        \gridLine{0.65}{-0.5}{1}{-0.5}
         \circle{-0}{-0.5}
        \gridLine{-0.15}{-0.5}{-0.5}{-0.5}
 \end{tikzpicture}~.
\end{eqnarray}
Upon using this we obtain
\begin{align}
    \begin{tikzpicture}[baseline={([yshift=0.6ex]current bounding box.center)},scale=0.45]
{\prop{2}{2}{colSt}{}}
   \foreach \x in {1,3}{\prop{\x}{1}{colSt}{}}   
    \foreach \x in {0,2,...,4}{\prop{\x}{0}{colSt}{}}
    \foreach \x in {-1,1,...,5}{\prop{\x}{-1}{colSt}{}}
    \foreach \x in {-2,0,...,6}{\prop{\x}{-2}{colSt}{}}
    \foreach \x in {-3,-1,...,5}{\prop{\x}{-3}{colSt}{}}
        \foreach \x in {7}{\newprop{\x}{-3}{colSt}{}}
    \foreach \x in {-4,-2,...,6}{\prop{\x}{-4}{colSt}{}}
     \foreach \x in {8}{\newprop{\x}{-4}{colSt}{}}
   \foreach \x in {-4,-2} {\propimp{6}{\x}{colimp}{}}
    \foreach \x in {-5.5,-3.5,...,4.5}{\istate{\x}{-4.5}{colSt}}
        \foreach \x in {6.5,8.5}{\newistate{\x}{-4.5}{colSt}}
    \foreach \x in {-4,...,1}{\tgridLine{\x-.5}{\x+2.5}{\x}{\x+2}
    \square{\x-.5}{\x+2.5}
    \circle{\x}{\x+2}}
     \foreach \x in {-6,...,1}{\circle{\x+0.5}{\x+1.5}}
    \foreach \x in {-6,...,1}{\circle
    {4-\x-0.5}{\x+1.5}}
      \draw[semithick,decorate,decoration={brace}] (-5,-1.15) -- (.25,4.15) node[midway,xshift=-7.5pt,yshift=7.5pt] {$\ell$};
      \draw[semithick,decorate,decoration={brace}] (-6.2,-4.45) -- (-4.65,-2.9) node[midway,xshift=-11.5pt,yshift=7.5pt] {$2t-\ell$};
\draw[semithick,decorate,decoration={brace}] (2.7,3.3) -- (6,-0.05) node[midway,xshift=7.5pt,yshift=7.5pt] {$2r$};
  \end{tikzpicture}
  \hspace{-.5cm}. 
\end{align}
This diagram can be fully contracted and, after properly normalizing, we obtain the $t>t_*$ part of Eq.~(21) of the main text.

\section{Entanglement evolution for random gates with maximal entangling power}\label{sec:maximal}
 In this section, we detail the exact calculation of the second R\'enyi entanglement entropy evolution for random dual unitary bulk gates with maximal entangling power, $p=1$, in the large $q$ limit. The entangling power of a bulk gate $U$ is defined as
\be
p=\frac{1}{q^2\left(q^2-1\right)}\left(q^4-\tr\left[\left(\tilde{U}^{{\rm T}_2}\left(\tilde{U}^{{\rm T}_2}\right)^{\dagger}\right)^2\right]\right),
\ee
where ${\rm T}_2$ represents the partial transpose with respect to the second qudit on which $\tilde{U}$ acts. Upon averaging, the bulk gate assumes a simple form, written in terms of $q$ and $p$. Indeed, with respect to the orthogonal basis $\left\{\begin{tikzpicture}[baseline={([yshift=-0.6ex]current bounding box.center)},scale=0.5]
    \circle{-0.5}{0.5}
      \gridLine{-0.38}{0.5}{0.2}{0.5}
 \end{tikzpicture},\begin{tikzpicture}[baseline={([yshift=-0.6ex]current bounding box.center)},scale=0.5]
    \circleblack{-0.5}{0.5}
      \gridLine{-0.38}{0.5}{0.2}{0.5}
 \end{tikzpicture}\right\}$ of the subspace onto which the gate is projected, it can be represented in the following matrix form (assuming $q\gg1$)~\cite{foligno2023growth}:
  \be\label{eq:AveragedGateMatrix}
\begin{aligned}
 &\begin{tikzpicture}[baseline={([yshift=-0.6ex]current bounding box.center)},scale=0.5]
   \prop{0}{0}{colSt}{}
 \end{tikzpicture}\approx\left(\begin{array}{cccc}
1 & 0 & 0 & 0\\
0 & 0 & 1-p & p/q\\
0 & 1-p & 0 & p/q\\
0 & p/q & p/q & 1
\end{array}\right).
\end{aligned}
\ee
We now substitute $p=1$ and proceed to prove Eq.~(21) of the main text.
We will use the fact that $\begin{tikzpicture}[baseline={([yshift=-0.6ex]current bounding box.center)},scale=0.5]
    \square{-0.5}{0.5}
      \gridLine{-0.38}{0.5}{0.2}{0.5}
 \end{tikzpicture}\approx\begin{tikzpicture}[baseline={([yshift=-0.6ex]current bounding box.center)},scale=0.5]
    \circleblack{-0.5}{0.5}
      \gridLine{-0.38}{0.5}{0.2}{0.5}
 \end{tikzpicture}+\begin{tikzpicture}[baseline={([yshift=-0.6ex]current bounding box.center)},scale=0.5]
    \circle{-0.5}{0.5}
      \gridLine{-0.38}{0.5}{0.2}{0.5}
 \end{tikzpicture}/q$ for $q\gg1$.
 
 Two main properties of the averaged gates that will be used in the calculation are
 
\begin{eqnarray}\label{eq:MaximalEP1stProperty}
\begin{tikzpicture}[baseline={([yshift=0]current bounding box.center)},scale=0.45] 
      \foreach \x in {0,...,5}{\prop{\x}{-\x}{colSt}{}}   
       \foreach \x in {-1,...,4}{\circle{\x+0.5}{-\x-1.5}}
 \circleblack
    {5.5}{-5.5}
      \draw[semithick,decorate,decoration={brace}] (4,-5.75) -- (-1.25,-0.5) node[midway,xshift=-20.5pt,yshift=0] {$m$};
  \end{tikzpicture}\approx \frac{1}{q^m} \begin{tikzpicture}[baseline={([yshift=-0.6ex]current bounding box.center)},scale=0.5]
  \tgridLine{0}{0}{-0.4}{0.4}
  \circleblack{0}{0}
  \end{tikzpicture}\otimes \Big[\begin{tikzpicture}[baseline={([yshift=-0.6ex]current bounding box.center)},scale=0.5]
  \tgridLine{0}{0}{0.4}{0.4}
  \circleblack{0}{0}
  \end{tikzpicture}\,\Big]^{\otimes m},
  \end{eqnarray}
which immediately stems from the form of the averaged gate in~\eqref{eq:AveragedGateMatrix} for $p=1$, and

  \begin{eqnarray}\label{eq:MaximalEP2ndProperty}
\begin{tikzpicture}[baseline={([yshift=0]current bounding box.center)},scale=0.45] 
      \foreach \x in {0,...,5}{\prop{\x}{-\x}{colSt}{}}   
       \foreach \x in {-1,...,4}{\circleblack{\x+0.5}{-\x-1.5}}
       \square{-.5}{.5}
      \draw[semithick,decorate,decoration={brace}] (4,-5.75) -- (-1.25,-0.5) node[midway,xshift=-20.5pt,yshift=0] {$m$};
  \end{tikzpicture}\approx \ \Big[\begin{tikzpicture}[baseline={([yshift=-0.6ex]current bounding box.center)},scale=0.5]
  \tgridLine{0}{0}{0.4}{0.4}
  \square{0}{0}
  \end{tikzpicture}\,\Big]^{\otimes m}\otimes\begin{tikzpicture}[baseline={([yshift=-0.6ex]current bounding box.center)},scale=0.5]
  \tgridLine{0}{0}{0.4}{-0.4}
  \square{0}{0}
  \end{tikzpicture}-\frac{1}{q^2}\Big[\begin{tikzpicture}[baseline={([yshift=-0.6ex]current bounding box.center)},scale=0.5]
  \tgridLine{0}{0}{0.4}{0.4}
  \square{0}{0}
  \end{tikzpicture}\,\Big]^{\otimes \left(m-1\right)}\otimes \begin{tikzpicture}[baseline={([yshift=-0.6ex]current bounding box.center)},scale=0.5]
  \tgridLine{0}{0}{0.4}{0.4}
  \circle{0}{0}
  \end{tikzpicture} \otimes \begin{tikzpicture}[baseline={([yshift=-0.6ex]current bounding box.center)},scale=0.5]
  \tgridLine{0}{0}{0.4}{-0.4}
  \circle{0}{0}
  \end{tikzpicture}\,,
  \end{eqnarray}
which holds to a leading order in $1/q$, and may be straightforwardly proved by induction. Eqs.~\eqref{eq:MaximalEP1stProperty} and~\eqref{eq:MaximalEP2ndProperty} mean that when the two sides of an equation are written in the basis of tensor products of $\left\{\begin{tikzpicture}[baseline={([yshift=-0.6ex]current bounding box.center)},scale=0.5]
    \circle{-0.5}{0.5}
      \gridLine{-0.38}{0.5}{0.2}{0.5}
 \end{tikzpicture},\begin{tikzpicture}[baseline={([yshift=-0.6ex]current bounding box.center)},scale=0.5]
    \circleblack{-0.5}{0.5}
      \gridLine{-0.38}{0.5}{0.2}{0.5}
 \end{tikzpicture}\right\}$, they feature the same coefficient multiplying each basis state, at the leading order in $1/q$.

We focus on the non-universal time regime $\ell/2<t<\ell+r$ (recall that outside this regime the results are independent of the bulk gates), and begin with the earlier regime $\ell/2<t\le\ell/2+r$. Applying the decomposition $\begin{tikzpicture}[baseline={([yshift=-0.6ex]current bounding box.center)},scale=0.65]
  \istate{0}{0}{colSt}
\end{tikzpicture}=(\begin{tikzpicture}[baseline={([yshift=-0.6ex]current bounding box.center)},scale=0.65]
   \tgridLine{0}{0}{0.35}{0.35}
  \circle{0}{0}
   \tgridLine{-.5}{0}{-.8}{0.3}
  \circle{-0.5}{0}
\end{tikzpicture}+\begin{tikzpicture}[baseline={([yshift=-0.6ex]current bounding box.center)},scale=0.65]
  \tgridLine{0}{0}{0.3}{0.3}
  \circleblack{0}{0}
  \tgridLine{-.5}{0}{-.8}{0.3}
  \circleblack{-0.5}{0}
  \end{tikzpicture})/q^4$ to the leftmost edge of the purity diagram, and using the properties in~\eqref{eq:MaximalEP1stProperty} and~\eqref{eq:MaximalEP2ndProperty}, we see that 
   \begin{eqnarray}\nonumber
\!\!\!\tr[\rho^2_A(t)]&=&\begin{tikzpicture}[baseline={([yshift=0.6ex]current bounding box.center)},scale=0.45]
{\prop{2}{2}{colSt}{}}
   \foreach \x in {1,3}{\prop{\x}{1}{colSt}{}}   
    \foreach \x in {0,2,...,4}{\prop{\x}{0}{colSt}{}}
    \foreach \x in {-1,1,...,5}{\prop{\x}{-1}{colSt}{}}
    \foreach \x in {-2,0,...,6}{\prop{\x}{-2}{colSt}{}}
    \foreach \x in {-3,-1,...,5}{\prop{\x}{-3}{colSt}{}}
    \foreach \x in {-2,0,...,6}{\prop{\x}{-4}{colSt}{}}
   \foreach \x in {-4,-2} {\prop{6}{\x}{colimp}{}}
    \foreach \x in {-1.5,0.5,...,4.5}{\istate{\x}{-4.5}{colSt}}
    \foreach \x in {-4,...,1}{\square{\x+0.5}{\x+1.5}}
	\circle{-3.5}{-3.5}
	\circle{-2.5}{-4.5}
    \foreach \x in {-3,...,1}{\circle
    {4-\x-0.5}{\x+1.5}}
     \foreach \x in {-6,...,-4}{\circle
    {6.5}{\x+1.5}}
      \draw[semithick,decorate,decoration={brace}] (-4,-2.25) -- (1.25,3) node[midway,xshift=-7.5pt,yshift=7.5pt] {$\ell$};
      \draw[semithick,decorate,decoration={brace}] (-2.75,-4.8) -- (-4.,-3.55) node[midway,xshift=-21.5pt] {$2t-\ell$};
  \draw[semithick,decorate,decoration={brace}] (2.55,2.85) -- (5.8,-0.3) node[midway,xshift=7.5pt,yshift=7.5pt] {$2r$};
  \end{tikzpicture}
\times \,q^{\ell+2r-6t}\\\nonumber
&\approx& \begin{tikzpicture}[baseline={([yshift=0.6ex]current bounding box.center)},scale=0.45]
{\prop{2}{2}{colSt}{}}
   \foreach \x in {1,3}{\prop{\x}{1}{colSt}{}}   
    \foreach \x in {0,2,...,4}{\prop{\x}{0}{colSt}{}}
    \foreach \x in {-1,1,...,5}{\prop{\x}{-1}{colSt}{}}
    \foreach \x in {-2,0,...,6}{\prop{\x}{-2}{colSt}{}}
    \foreach \x in {-1,1,...,5}{\prop{\x}{-3}{colSt}{}}
    \foreach \x in {0,2,...,6}{\prop{\x}{-4}{colSt}{}}
   \foreach \x in {-4,-2} {\prop{6}{\x}{colimp}{}}
    \foreach \x in {0.5,2.5,4.5}{\istate{\x}{-4.5}{colSt}}
    \foreach \x in {-3,...,1}{\square{\x+0.5}{\x+1.5}}
\circle{-2.5}{-2.5}
	\circle{-1.5}{-3.5}
	\circle{-0.5}{-4.5}
    \foreach \x in {-3,...,1}{\circle
    {4-\x-0.5}{\x+1.5}}
     \foreach \x in {-6,...,-4}{\circle
    {6.5}{\x+1.5}}
          \draw[semithick,decorate,decoration={brace}] (-3,-1.25) -- (1.25,3) node[midway,xshift=-10.5pt,yshift=7.5pt] {$\ell-1$};
  \end{tikzpicture}
\times \,q^{\ell+2r-6t-3}+\begin{tikzpicture}[baseline={([yshift=0.6ex]current bounding box.center)},scale=0.45]
{\prop{2}{2}{colSt}{}}
   \foreach \x in {1,3}{\prop{\x}{1}{colSt}{}}   
    \foreach \x in {0,2,...,4}{\prop{\x}{0}{colSt}{}}
    \foreach \x in {-1,1,...,5}{\prop{\x}{-1}{colSt}{}}
    \foreach \x in {0,2,...,6}{\prop{\x}{-2}{colSt}{}}
    \foreach \x in {1,3,5}{\prop{\x}{-3}{colSt}{}}
    \foreach \x in {2,4,6}{\prop{\x}{-4}{colSt}{}}
   \foreach \x in {-4,-2} {\prop{6}{\x}{colimp}{}}
    \foreach \x in {2.5,4.5}{\istate{\x}{-4.5}{colSt}}
    \foreach \x in {-2,...,1}{\square{\x+0.5}{\x+1.5}}
\foreach \x in {0,...,3}  {\square{-1.5+\x}{-1.5-\x}}
    \foreach \x in {-3,...,1}{\circle
    {4-\x-0.5}{\x+1.5}}
     \foreach \x in {-6,...,-4}{\circle
    {6.5}{\x+1.5}}
              \draw[semithick,decorate,decoration={brace}] (-2,-0.25) -- (1.25,3) node[midway,xshift=-10.5pt,yshift=7.5pt] {$\ell-2$};
  \end{tikzpicture}
\times \,q^{2\ell+2r-8t-4}\\
&&-\begin{tikzpicture}[baseline={([yshift=0.6ex]current bounding box.center)},scale=0.45]
{\prop{2}{2}{colSt}{}}
   \foreach \x in {1,3}{\prop{\x}{1}{colSt}{}}   
    \foreach \x in {0,2,...,4}{\prop{\x}{0}{colSt}{}}
    \foreach \x in {-1,1,...,5}{\prop{\x}{-1}{colSt}{}}
    \foreach \x in {0,2,...,6}{\prop{\x}{-2}{colSt}{}}
    \foreach \x in {1,3,5}{\prop{\x}{-3}{colSt}{}}
    \foreach \x in {2,4,6}{\prop{\x}{-4}{colSt}{}}
   \foreach \x in {-4,-2} {\prop{6}{\x}{colimp}{}}
    \foreach \x in {2.5,4.5}{\istate{\x}{-4.5}{colSt}}
    \foreach \x in {-2,...,1}{\square{\x+0.5}{\x+1.5}}
\foreach \x in {0,...,1}  {\square{-1.5+\x}{-1.5-\x}}
\foreach \x in {2,...,3}  {\circle{-1.5+\x}{-1.5-\x}}
    \foreach \x in {-3,...,1}{\circle
    {4-\x-0.5}{\x+1.5}}
     \foreach \x in {-6,...,-4}{\circle
    {6.5}{\x+1.5}}
       \draw[semithick,decorate,decoration={brace}] (-2,-0.25) -- (1.25,3) node[midway,xshift=-10.5pt,yshift=7.5pt] {$\ell-2$};
             \draw[semithick,decorate,decoration={brace}] (-0.75,-2.8) -- (-2.,-1.55) node[midway,xshift=-21.5pt] {$2t-\ell$};
  \end{tikzpicture}\times \,q^{2\ell+2r-8t-6}\,,
  \end{eqnarray}
  which reduces to
  \begin{eqnarray}\nonumber
\tr[\rho^2_A(t)]&\approx& \begin{tikzpicture}[baseline={([yshift=0.6ex]current bounding box.center)},scale=0.45]
{\prop{2}{2}{colSt}{}}
   \foreach \x in {1,3}{\prop{\x}{1}{colSt}{}}   
    \foreach \x in {0,2,...,4}{\prop{\x}{0}{colSt}{}}
    \foreach \x in {-1,1,...,5}{\prop{\x}{-1}{colSt}{}}
    \foreach \x in {-2,0,...,6}{\prop{\x}{-2}{colSt}{}}
    \foreach \x in {-1,1,...,5}{\prop{\x}{-3}{colSt}{}}
    \foreach \x in {0,2,...,6}{\prop{\x}{-4}{colSt}{}}
   \foreach \x in {-4,-2} {\prop{6}{\x}{colimp}{}}
    \foreach \x in {0.5,2.5,4.5}{\istate{\x}{-4.5}{colSt}}
    \foreach \x in {-3,...,1}{\square{\x+0.5}{\x+1.5}}
\circle{-2.5}{-2.5}
	\circle{-1.5}{-3.5}
	\circle{-0.5}{-4.5}
    \foreach \x in {-3,...,1}{\circle
    {4-\x-0.5}{\x+1.5}}
     \foreach \x in {-6,...,-4}{\circle
    {6.5}{\x+1.5}}
             \draw[semithick,decorate,decoration={brace}] (-3,-1.25) -- (1.25,3) node[midway,xshift=-10.5pt,yshift=7.5pt] {$\ell-1$};
                          \draw[semithick,decorate,decoration={brace}] (-0.75,-4.8) -- (-3.,-2.55) node[midway,xshift=-28.5pt] {$2t-\ell+1$};
  \end{tikzpicture}
\times \,q^{\ell+2r-6t-3}+\left(\,\begin{tikzpicture}[baseline={([yshift=-0.6ex]current bounding box.center)},scale=0.5]
   \prop{1}{0}{colimp}{}
   \circle{1.5}{0.5}
    \square{0.5}{0.5}
       \circle{1.5}{-0.5}
    \square{0.5}{-0.5}
 \end{tikzpicture}\,\right)^{t-r}
\times \,q^{2r-6t}\\
&&-\begin{tikzpicture}[baseline={([yshift=0.6ex]current bounding box.center)},scale=0.45]
    \foreach \x in {4}{\prop{\x}{0}{colSt}{}}
    \foreach \x in {3,5}{\prop{\x}{-1}{colSt}{}}
    \foreach \x in {2,4,6}{\prop{\x}{-2}{colSt}{}}
    \foreach \x in {1,3,5}{\prop{\x}{-3}{colSt}{}}
    \foreach \x in {2,4,6}{\prop{\x}{-4}{colSt}{}}
   \foreach \x in {-4,-2} {\prop{6}{\x}{colimp}{}}
    \foreach \x in {2.5,4.5}{\istate{\x}{-4.5}{colSt}}
     \foreach \x in {-6,...,-4}{\circle
    {6.5}{\x+1.5}}
     \foreach \x in {-2,...,1}{\square{\x+2.5}{\x-0.5}}
\foreach \x in {2,3}  {\circle{-1.5+\x}{-1.5-\x}}
\foreach \x in {2,3,4}  {\circle{2.5+\x}{2.5-\x}}
       \draw[semithick,decorate,decoration={brace}] (-0,-2) -- (3.25,1) node[midway,xshift=-10.5pt,yshift=7.5pt] {$\ell-2$};
             \draw[semithick,decorate,decoration={brace}] (1.25,-4.8) -- (0.,-3.55) node[midway,xshift=-12.5pt] {$2$};
               \draw[semithick,decorate,decoration={brace}] (7,-1.25) -- (7,-4.75) node[midway,xshift=21.5pt] {$2t-2r$};
  \end{tikzpicture}\times \,q^{\ell+2r-6t-6}\,.\label{eq:PurityEarlyTimeMaximalEP}
  \end{eqnarray}
We now show that the last term in~\eqref{eq:PurityEarlyTimeMaximalEP} (with the negative sign) is subleading in $1/q$ compared to the first two. Indeed, in this last term, we resolve the initial state at the leftmost edge of the diagram as before, and upper-bound  
the contribution of the diagram with $\begin{tikzpicture}[baseline={([yshift=-0.6ex]current bounding box.center)},scale=0.5]
    \circleblack{-0.5}{0.5}
      \gridLine{-0.38}{0.5}{0.2}{0.5}
 \end{tikzpicture}$ by replacing these states with $\begin{tikzpicture}[baseline={([yshift=-0.6ex]current bounding box.center)},scale=0.5]
    \square{-0.5}{0.5}
      \gridLine{-0.38}{0.5}{0.2}{0.5}
 \end{tikzpicture}$. By doing so repeatedly until the diagram is completely reduced, we find that
  \begin{eqnarray}
\begin{tikzpicture}[baseline={([yshift=0.6ex]current bounding box.center)},scale=0.45]
    \foreach \x in {4}{\prop{\x}{0}{colSt}{}}
    \foreach \x in {3,5}{\prop{\x}{-1}{colSt}{}}
    \foreach \x in {2,4,6}{\prop{\x}{-2}{colSt}{}}
    \foreach \x in {1,3,5}{\prop{\x}{-3}{colSt}{}}
    \foreach \x in {2,4,6}{\prop{\x}{-4}{colSt}{}}
   \foreach \x in {-4,-2} {\prop{6}{\x}{colimp}{}}
    \foreach \x in {2.5,4.5}{\istate{\x}{-4.5}{colSt}}
     \foreach \x in {-6,...,-4}{\circle
    {6.5}{\x+1.5}}
     \foreach \x in {-2,...,1}{\square{\x+2.5}{\x-0.5}}
\foreach \x in {2,3}  {\circle{-1.5+\x}{-1.5-\x}}
\foreach \x in {2,3,4}  {\circle{2.5+\x}{2.5-\x}}
       \draw[semithick,decorate,decoration={brace}] (-0,-2) -- (3.25,1) node[midway,xshift=-10.5pt,yshift=7.5pt] {$\ell-2$};
             \draw[semithick,decorate,decoration={brace}] (1.25,-4.8) -- (0.,-3.55) node[midway,xshift=-12.5pt] {$2$};
               \draw[semithick,decorate,decoration={brace}] (7,-1.25) -- (7,-4.75) node[midway,xshift=21.5pt] {$2t-2r$};
  \end{tikzpicture}\times \,q^{\ell+2r-6t-6} \leq \frac{1}{q^{4t}}+\left(\,\begin{tikzpicture}[baseline={([yshift=-0.6ex]current bounding box.center)},scale=0.5]
   \prop{1}{0}{colimp}{}
   \circle{1.5}{0.5}
    \square{0.5}{0.5}
       \circle{1.5}{-0.5}
    \square{0.5}{-0.5}
 \end{tikzpicture}\,\right)^{t-r}\times \,q^{2r-6t-4}.
  \end{eqnarray}
In contrast, the first term in~\eqref{eq:PurityEarlyTimeMaximalEP} can be lower-bounded as
\be
\begin{tikzpicture}[baseline={([yshift=0.6ex]current bounding box.center)},scale=0.45]
{\prop{2}{2}{colSt}{}}
   \foreach \x in {1,3}{\prop{\x}{1}{colSt}{}}   
    \foreach \x in {0,2,...,4}{\prop{\x}{0}{colSt}{}}
    \foreach \x in {-1,1,...,5}{\prop{\x}{-1}{colSt}{}}
    \foreach \x in {-2,0,...,6}{\prop{\x}{-2}{colSt}{}}
    \foreach \x in {-1,1,...,5}{\prop{\x}{-3}{colSt}{}}
    \foreach \x in {0,2,...,6}{\prop{\x}{-4}{colSt}{}}
   \foreach \x in {-4,-2} {\prop{6}{\x}{colimp}{}}
    \foreach \x in {0.5,2.5,4.5}{\istate{\x}{-4.5}{colSt}}
    \foreach \x in {-3,...,1}{\square{\x+0.5}{\x+1.5}}
\circle{-2.5}{-2.5}
	\circle{-1.5}{-3.5}
	\circle{-0.5}{-4.5}
    \foreach \x in {-3,...,1}{\circle
    {4-\x-0.5}{\x+1.5}}
     \foreach \x in {-6,...,-4}{\circle
    {6.5}{\x+1.5}}
             \draw[semithick,decorate,decoration={brace}] (-3,-1.25) -- (1.25,3) node[midway,xshift=-10.5pt,yshift=7.5pt] {$\ell-1$};
                          \draw[semithick,decorate,decoration={brace}] (-0.75,-4.8) -- (-3.,-2.55) node[midway,xshift=-28.5pt] {$2t-\ell+1$};
  \end{tikzpicture}\times \,q^{\ell+2r-6t-3}\ge \frac{1}{q^{2\ell}},
\ee
by resolving the initial state and keeping only the contributions of $\begin{tikzpicture}[baseline={([yshift=-0.6ex]current bounding box.center)},scale=0.5]
    \circle{-0.5}{0.5}
      \gridLine{-0.38}{0.5}{0.2}{0.5}
 \end{tikzpicture}$ states. Given that $t>\ell/2$, the last term of~\eqref{eq:PurityEarlyTimeMaximalEP} is thus upper-bounded by an expression that is subleading in $1/q$ compared to the first two terms, meaning that we may omit it and write
 \begin{eqnarray}
  \!\!\!\tr[\rho^2_A(t)]&\approx&
\begin{tikzpicture}[baseline={([yshift=0.6ex]current bounding box.center)},scale=0.45]
{\prop{2}{2}{colSt}{}}
   \foreach \x in {1,3}{\prop{\x}{1}{colSt}{}}   
    \foreach \x in {0,2,...,4}{\prop{\x}{0}{colSt}{}}
    \foreach \x in {-1,1,...,5}{\prop{\x}{-1}{colSt}{}}
    \foreach \x in {-2,0,...,6}{\prop{\x}{-2}{colSt}{}}
    \foreach \x in {-1,1,...,5}{\prop{\x}{-3}{colSt}{}}
    \foreach \x in {0,2,...,6}{\prop{\x}{-4}{colSt}{}}
   \foreach \x in {-4,-2} {\prop{6}{\x}{colimp}{}}
    \foreach \x in {0.5,2.5,4.5}{\istate{\x}{-4.5}{colSt}}
    \foreach \x in {-3,...,1}{\square{\x+0.5}{\x+1.5}}
\circle{-2.5}{-2.5}
	\circle{-1.5}{-3.5}
	\circle{-0.5}{-4.5}
    \foreach \x in {-3,...,1}{\circle
    {4-\x-0.5}{\x+1.5}}
     \foreach \x in {-6,...,-4}{\circle
    {6.5}{\x+1.5}}
             \draw[semithick,decorate,decoration={brace}] (-3,-1.25) -- (1.25,3) node[midway,xshift=-10.5pt,yshift=7.5pt] {$\ell-1$};
                          \draw[semithick,decorate,decoration={brace}] (-0.75,-4.8) -- (-3.,-2.55) node[midway,xshift=-28.5pt] {$2t-\ell+1$};
  \end{tikzpicture}\times \frac{1}{q^{6t-\ell-2r+3}}+\frac{1}{q^{6t-2r}}\left(\,\begin{tikzpicture}[baseline={([yshift=-0.6ex]current bounding box.center)},scale=0.5]
   \prop{1}{0}{colimp}{}
   \circle{1.5}{0.5}
    \square{0.5}{0.5}
       \circle{1.5}{-0.5}
    \square{0.5}{-0.5}
 \end{tikzpicture}\,\right)^{t-r}\,.
  \end{eqnarray}
Finally, we may apply this rule repeatedly to reduce the remaining diagram, obtaining
\be\label{eq:ExactPurityMaximalEP}
\tr[\rho^2_A(t)]\approx\frac{1}{q^{2\ell}}+\frac{1}{q^{2t+2r}}e^{\left(r-t\right)S_{\rm imp}^{\left(2\right)}}.
\ee
An analogous treatment of the later non-universal time regime, namely $\ell/2+r<t<\ell+r$, yields the exact same form of the purity as in~\eqref{eq:ExactPurityMaximalEP}.

The result of~\eqref{eq:ExactPurityMaximalEP} should be interpreted like so: the exact expression for the purity contains two terms, but almost for any $t$ one of the terms dominates the other to a leading order in $1/q$, such that the contribution of the subleading term of the two may be neglected. We can define the time $t_*$ as that at which the two terms are comparable,
\be
\frac{1}{q^{2\ell}}\approx\frac{1}{q^{2t_*+2r}}e^{\left(r-t_*\right)S_{\rm imp}^{\left(2\right)}},
\ee
and then the evolution of the second R\'enyi entropy is given by Eq.~(21) of the main text.

\section{Proof of bounds in the non-universal time regime}\label{sec:bounds}
In this section, we prove two lower bounds to the purity of $\rho_A\!\left(t\right)$ in the non-universal time regime $\ell/2<t<\ell+r$, for averaged random dual unitary bulk gates with a uniform entangling power $p<1$ and assuming a large local Hilbert dimension, $q\gg\left(1-p\right)^{-1}$. The first bound, appearing in Eq.~(22) of the main text,
applies in the earlier time regime $\ell/2<t\le\ell/2+r$, while the second bound, appearing in Eq.~(23) of the main text,
applies in the later time regime $\ell/2+r<t<\ell+r$. In the derivation that follows, we use the matrix form of the averaged gate given in~\eqref{eq:AveragedGateMatrix}.


In order to prove the first bound,
we first note that, for any integer $m$,
\be\label{eq:RowDiagramEmptyCircles}
\begin{tikzpicture}[baseline={([yshift=-2ex]current bounding box.center)},scale=0.45] 
      \foreach \x in {-4,...,1}{\prop{\x+1}{\x+1}{colSt}{}}   
       \foreach \x in {-4,...,1}{\square{\x+0.5}{\x+1.5}}
 \circle
    {2.5}{2.5}
     \circle
    {-3.5}{-3.5}
      \draw[semithick,decorate,decoration={brace}] (-4,-2.25) -- (1.25,3) node[midway,xshift=-7.5pt,yshift=7.5pt] {$m$};
  \end{tikzpicture}\approx q^2 \Big[(1-p)\,\begin{tikzpicture}[baseline={([yshift=-0.6ex]current bounding box.center)},scale=0.5]
  \tgridLine{0}{0}{0.4}{-0.4}
  \circleblack{0}{0}
  \end{tikzpicture}+\frac{1}{q}\,\begin{tikzpicture}[baseline={([yshift=-0.6ex]current bounding box.center)},scale=0.5]
  \tgridLine{0}{0}{0.4}{-0.4}
  \circle{0}{0}
  \end{tikzpicture}\,\Big]^{\otimes m}.
\ee
This can be seen by writing $\begin{tikzpicture}[baseline={([yshift=-0.6ex]current bounding box.center)},scale=0.5]
    \square{-0.5}{0.5}
      \gridLine{-0.38}{0.5}{0.2}{0.5}
 \end{tikzpicture}\approx\begin{tikzpicture}[baseline={([yshift=-0.6ex]current bounding box.center)},scale=0.5]
    \circleblack{-0.5}{0.5}
      \gridLine{-0.38}{0.5}{0.2}{0.5}
 \end{tikzpicture}+\begin{tikzpicture}[baseline={([yshift=-0.6ex]current bounding box.center)},scale=0.5]
    \circle{-0.5}{0.5}
      \gridLine{-0.38}{0.5}{0.2}{0.5}
 \end{tikzpicture}/q$, and noting that whenever the operation $\begin{tikzpicture}[baseline={([yshift=-0.6ex]current bounding box.center)},scale=0.5]
   \prop{0}{0}{colSt}{}
   \circleblack{-0.5}{-0.5}
    \circle{-0.5}{0.5}
 \end{tikzpicture}\approx(1-p)\, 
\begin{tikzpicture}[baseline={([yshift=-0.6ex]current bounding box.center)},scale=0.5]
  \circle{-0.5}{-0.5}
    \circleblack{-0.5}{0.5}
    \gridLine{-0.38}{-0.5}{0.3}{-0.5}
      \gridLine{-0.38}{0.5}{0.3}{0.5}
 \end{tikzpicture}+p/q\begin{tikzpicture}[baseline={([yshift=-0.6ex]current bounding box.center)},scale=0.5]
  \circleblack{-0.5}{-0.5}
    \circleblack{-0.5}{0.5}
    \gridLine{-0.38}{-0.5}{0.3}{-0.5}
      \gridLine{-0.38}{0.5}{0.3}{0.5}
 \end{tikzpicture}$ is encountered while reducing the diagram, the second term on the right-hand side can be omitted as it produces only a subleading contribution in $1/q$. From~\eqref{eq:RowDiagramEmptyCircles} one can readily see that
 \be\label{eq:RowDiagramEmptyCirclesInequality}
\begin{tikzpicture}[baseline={([yshift=-2ex]current bounding box.center)},scale=0.45] 
      \foreach \x in {-4,...,1}{\prop{\x+1}{\x+1}{colSt}{}}   
       \foreach \x in {-4,...,1}{\square{\x+0.5}{\x+1.5}}
 \circle
    {2.5}{2.5}
     \circle
    {-3.5}{-3.5}
      \draw[semithick,decorate,decoration={brace}] (-4,-2.25) -- (1.25,3) node[midway,xshift=-7.5pt,yshift=7.5pt] {$m$};
  \end{tikzpicture}\geq q^2(1-p)^m \,\big[\,\begin{tikzpicture}[baseline={([yshift=-0.6ex]current bounding box.center)},scale=0.5]
  \tgridLine{0}{0}{0.4}{-0.4}
  \square{0}{0}
  \end{tikzpicture}\,\big]^{\otimes m}.
\ee
The inequality~\eqref{eq:RowDiagramEmptyCirclesInequality} should be interpreted so that when its two sides are contracted with a diagram with $m$ free legs that is built from the averaged dual-unitary gates in~\eqref{eq:AveragedGateMatrix} and the impurity gates, then the contraction with the right hand side yields a result not larger than that of the contraction with the left hand side. This observation relies on the fact that, due to the non-negative entries of~\eqref{eq:AveragedGateMatrix}, this contraction yields a $1/q$ expansion including only non-negative terms, so that omitting a term cannot increase the overall value of the contracted diagram.

Next, we apply~\eqref{eq:RowDiagramEmptyCirclesInequality} to the diagram that stands for the purity at a time $\ell/2<t\le\ell/2+r$, and observe that
 \begin{eqnarray}\nonumber
\!\!\!\tr[\rho^2_A(t)]&=&\begin{tikzpicture}[baseline={([yshift=0.6ex]current bounding box.center)},scale=0.45]
{\prop{2}{2}{colSt}{}}
   \foreach \x in {1,3}{\prop{\x}{1}{colSt}{}}   
    \foreach \x in {0,2,...,4}{\prop{\x}{0}{colSt}{}}
    \foreach \x in {-1,1,...,5}{\prop{\x}{-1}{colSt}{}}
    \foreach \x in {-2,0,...,6}{\prop{\x}{-2}{colSt}{}}
    \foreach \x in {-3,-1,...,5}{\prop{\x}{-3}{colSt}{}}
    \foreach \x in {-2,0,...,6}{\prop{\x}{-4}{colSt}{}}
   \foreach \x in {-4,-2} {\prop{6}{\x}{colimp}{}}
    \foreach \x in {-1.5,0.5,...,4.5}{\istate{\x}{-4.5}{colSt}}
    \foreach \x in {-4,...,1}{\square{\x+0.5}{\x+1.5}}
	\circle{-3.5}{-3.5}
	\circle{-2.5}{-4.5}
    \foreach \x in {-3,...,1}{\circle
    {4-\x-0.5}{\x+1.5}}
     \foreach \x in {-6,...,-4}{\circle
    {6.5}{\x+1.5}}
      \draw[semithick,decorate,decoration={brace}] (-4,-2.25) -- (1.25,3) node[midway,xshift=-7.5pt,yshift=7.5pt] {$\ell$};
      \draw[semithick,decorate,decoration={brace}] (-2.75,-4.8) -- (-4.,-3.55) node[midway,xshift=-21.5pt] {$2t-\ell$};
  \draw[semithick,decorate,decoration={brace}] (2.55,2.85) -- (5.8,-0.3) node[midway,xshift=7.5pt,yshift=7.5pt] {$2r$};
  \end{tikzpicture}
\times \,q^{\ell+2r-6t}\geq  \begin{tikzpicture}[baseline={([yshift=0]current bounding box.center)},scale=0.45]
    \foreach \x in {4}{\prop{\x}{0}{colSt}{}}
    \foreach \x in {3,5}{\prop{\x}{-1}{colSt}{}}
    \foreach \x in {2,4,6}{\prop{\x}{-2}{colSt}{}}
    \foreach \x in {1,3,5}{\prop{\x}{-3}{colSt}{}}
    \foreach \x in {0,2,...,6}{\prop{\x}{-4}{colSt}{}}
   \foreach \x in {-4,-2} {\prop{6}{\x}{colimp}{}}
    \foreach \x in {-1.5,0.5,...,4.5}{\istate{\x}{-4.5}{colSt}}
    \foreach \x in {-4,...,1}{\square{\x+2.5}{\x-0.5}}
    \foreach \x in {-3,...,-1}{\circle
    {4-\x-0.5}{\x+1.5}}
     \foreach \x in {-6,...,-4}{\circle
    {6.5}{\x+1.5}}
   \draw[semithick,decorate,decoration={brace}] (-2,-4.25) -- (3.25,1) node[midway,xshift=-7.5pt,yshift=7.5pt] {$\ell$};
  \draw[semithick,decorate,decoration={brace}] (4.55,1) -- (6,-0.35) node[midway,xshift=22.5pt,yshift=7.5pt] {$2r+\ell-2t$};
  \end{tikzpicture}\hspace{-1cm}  \times \,\frac{(1-p)^{\ell(2t-\ell)}}{q^{\ell+2t-2r}}
\\\label{eq:EarlyLowerBoundProof}
&=&\frac{(1-p)^{\ell(2t-\ell)}}{q^{2\ell+2t-2r}}\left(\,\begin{tikzpicture}[baseline={([yshift=-0.6ex]current bounding box.center)},scale=0.5]
   \prop{1}{0}{colimp}{}
   \circle{1.5}{0.5}
    \square{0.5}{0.5}
       \circle{1.5}{-0.5}
    \square{0.5}{-0.5}
 \end{tikzpicture}\,\right)^{t-r}\,,
  \end{eqnarray}
which proves the first bound, given in Eq.~(22) of the main text.

To prove the second bound,
we will repeatedly use the decomposition $\begin{tikzpicture}[baseline={([yshift=-0.6ex]current bounding box.center)},scale=0.65]
  \istate{0}{0}{colSt}
\end{tikzpicture}=(\begin{tikzpicture}[baseline={([yshift=-0.6ex]current bounding box.center)},scale=0.65]
   \tgridLine{0}{0}{0.35}{0.35}
  \circle{0}{0}
   \tgridLine{-.5}{0}{-.8}{0.3}
  \circle{-0.5}{0}
\end{tikzpicture}+\begin{tikzpicture}[baseline={([yshift=-0.6ex]current bounding box.center)},scale=0.65]
  \tgridLine{0}{0}{0.3}{0.3}
  \circleblack{0}{0}
  \tgridLine{-.5}{0}{-.8}{0.3}
  \circleblack{-0.5}{0}
  \end{tikzpicture})/q^4$ of the initial state to reduce the purity diagram for $\ell/2+r<t<\ell+r$. The first step leads to
  \begin{eqnarray}\nonumber
\!\!\!\tr[\rho^2_A(t)]&=&\begin{tikzpicture}[baseline={([yshift=-4ex]current bounding box.center)},scale=0.45]
{\prop{2}{2}{colSt}{}}
   \foreach \x in {1,3}{\prop{\x}{1}{colSt}{}}   
    \foreach \x in {0,2,...,4}{\prop{\x}{0}{colSt}{}}
    \foreach \x in {-1,1,...,5}{\prop{\x}{-1}{colSt}{}}
    \foreach \x in {-2,0,...,6}{\prop{\x}{-2}{colSt}{}}
    \foreach \x in {-3,-1,...,5}{\prop{\x}{-3}{colSt}{}}
        \foreach \x in {-1,1,...,5}{\prop{\x}{-5}{colSt}{}}
                \foreach \x in {1,3,5}{\prop{\x}{-7}{colSt}{}}
    \foreach \x in {-2,0,...,6}{\prop{\x}{-4}{colSt}{}}
        \foreach \x in {0,2,...,6}{\prop{\x}{-6}{colSt}{}}
                \foreach \x in {2,4}{\prop{\x}{-8}{colSt}{}}
   \foreach \x in {-8,-6,-4,-2} {\prop{6}{\x}{colimp}{}}
    \foreach \x in {2.5,4.5}{\istate{\x}{-8.5}{colSt}}
    \foreach \x in {-4,...,1}{\square{\x+0.5}{\x+1.5}}
	\circle{-3.5}{-3.5}
	\circle{-2.5}{-4.5}
	    \foreach \x in {1,...,6}{\circle
    {-4.5+\x}{-2.5-\x}}
    \foreach \x in {-3,...,1}{\circle
    {4-\x-0.5}{\x+1.5}}
     \foreach \x in {-10,...,-4}{\circle
    {6.5}{\x+1.5}}
      \draw[semithick,decorate,decoration={brace}] (-4,-2.25) -- (1.25,3) node[midway,xshift=-7.5pt,yshift=7.5pt] {$\ell$};
      \draw[semithick,decorate,decoration={brace}] (1.25,-8.8) -- (-4.,-3.55) node[midway,xshift=-21.5pt] {$2t-\ell$};
  \draw[semithick,decorate,decoration={brace}] (2.55,2.85) -- (5.8,-0.3) node[midway,xshift=7.5pt,yshift=7.5pt] {$2r$};
  \end{tikzpicture}
\times \,q^{\ell+2r-6t}\\\nonumber
&=&\begin{tikzpicture}[baseline={([yshift=-4ex]current bounding box.center)},scale=0.45]
{\prop{2}{2}{colSt}{}}
   \foreach \x in {1,3}{\prop{\x}{1}{colSt}{}}   
    \foreach \x in {0,2,...,4}{\prop{\x}{0}{colSt}{}}
    \foreach \x in {-1,1,...,5}{\prop{\x}{-1}{colSt}{}}
    \foreach \x in {-2,0,...,6}{\prop{\x}{-2}{colSt}{}}
    \foreach \x in {-3,-1,...,5}{\prop{\x}{-3}{colSt}{}}
        \foreach \x in {-1,1,...,5}{\prop{\x}{-5}{colSt}{}}
                \foreach \x in {1,3,5}{\prop{\x}{-7}{colSt}{}}
    \foreach \x in {-2,0,...,6}{\prop{\x}{-4}{colSt}{}}
        \foreach \x in {0,2,...,6}{\prop{\x}{-6}{colSt}{}}
                \foreach \x in {2,4}{\prop{\x}{-8}{colSt}{}}
   \foreach \x in {-8,-6,-4,-2} {\prop{6}{\x}{colimp}{}}
    \foreach \x in {4.5}{\istate{\x}{-8.5}{colSt}}
    \foreach \x in {-4,...,1}{\square{\x+0.5}{\x+1.5}}
	    \foreach \x in {1,...,6}{\circle
    {-4.5+\x}{-2.5-\x}}
    \foreach \x in {-3,...,1}{\circle
    {4-\x-0.5}{\x+1.5}}
     \foreach \x in {-10,...,-4}{\circle
    {6.5}{\x+1.5}}
    \circle{2.5}{-8.5}
     \circle{3.5}{-8.5}
      \draw[semithick,decorate,decoration={brace}] (-4,-2.25) -- (1.25,3) node[midway,xshift=-7.5pt,yshift=7.5pt] {$\ell$};
      \draw[semithick,decorate,decoration={brace}] (1.25,-8.8) -- (-4.,-3.55) node[midway,xshift=-21.5pt] {$2t-\ell$};
  \draw[semithick,decorate,decoration={brace}] (2.55,2.85) -- (5.8,-0.3) node[midway,xshift=7.5pt,yshift=7.5pt] {$2r$};
  \end{tikzpicture}
\times \,q^{\ell+2r-6t-4}+\begin{tikzpicture}[baseline={([yshift=-4ex]current bounding box.center)},scale=0.45]
{\prop{2}{2}{colSt}{}}
   \foreach \x in {1,3}{\prop{\x}{1}{colSt}{}}   
    \foreach \x in {0,2,...,4}{\prop{\x}{0}{colSt}{}}
    \foreach \x in {-1,1,...,5}{\prop{\x}{-1}{colSt}{}}
    \foreach \x in {-2,0,...,6}{\prop{\x}{-2}{colSt}{}}
    \foreach \x in {-3,-1,...,5}{\prop{\x}{-3}{colSt}{}}
        \foreach \x in {-1,1,...,5}{\prop{\x}{-5}{colSt}{}}
                \foreach \x in {1,3,5}{\prop{\x}{-7}{colSt}{}}
    \foreach \x in {-2,0,...,6}{\prop{\x}{-4}{colSt}{}}
        \foreach \x in {0,2,...,6}{\prop{\x}{-6}{colSt}{}}
                \foreach \x in {2,4}{\prop{\x}{-8}{colSt}{}}
   \foreach \x in {-8,-6,-4,-2} {\prop{6}{\x}{colimp}{}}
    \foreach \x in {4.5}{\istate{\x}{-8.5}{colSt}}
    \foreach \x in {-4,...,1}{\square{\x+0.5}{\x+1.5}}
	    \foreach \x in {1,...,6}{\circle
    {-4.5+\x}{-2.5-\x}}
    \foreach \x in {-3,...,1}{\circle
    {4-\x-0.5}{\x+1.5}}
     \foreach \x in {-10,...,-4}{\circle
    {6.5}{\x+1.5}}
        \circleblack{2.5}{-8.5}
     \circleblack{3.5}{-8.5}
  \end{tikzpicture}
\times \,q^{\ell+2r-6t-4},\\
\end{eqnarray}
which, by using the fact that $\begin{tikzpicture}[baseline={([yshift=-0.6ex]current bounding box.center)},scale=0.5]
   \prop{0}{0}{colSt}{}
   \circleblack{-0.5}{-0.5}
    \circle{-0.5}{0.5}
 \end{tikzpicture}\ge(1-p)\, 
\begin{tikzpicture}[baseline={([yshift=-0.6ex]current bounding box.center)},scale=0.5]
  \circle{-0.5}{-0.5}
    \circleblack{-0.5}{0.5}
    \gridLine{-0.38}{-0.5}{0.3}{-0.5}
      \gridLine{-0.38}{0.5}{0.3}{0.5}
 \end{tikzpicture}\,$, yields
\be
\!\!\!\tr[\rho^2_A(t)]\geq\begin{tikzpicture}[baseline={([yshift=-4ex]current bounding box.center)},scale=0.45]
{\prop{2}{2}{colSt}{}}
   \foreach \x in {1,3}{\prop{\x}{1}{colSt}{}}   
    \foreach \x in {0,2,...,4}{\prop{\x}{0}{colSt}{}}
    \foreach \x in {-1,1,...,5}{\prop{\x}{-1}{colSt}{}}
    \foreach \x in {-2,0,...,6}{\prop{\x}{-2}{colSt}{}}
    \foreach \x in {-1,1,...,5}{\prop{\x}{-3}{colSt}{}}
        \foreach \x in {1,3,5}{\prop{\x}{-5}{colSt}{}}
                \foreach \x in {3,5}{\prop{\x}{-7}{colSt}{}}
    \foreach \x in {0,2,...,6}{\prop{\x}{-4}{colSt}{}}
        \foreach \x in {2,4,6}{\prop{\x}{-6}{colSt}{}}
                \foreach \x in {4}{\prop{\x}{-8}{colSt}{}}
   \foreach \x in {-8,-6,-4,-2} {\prop{6}{\x}{colimp}{}}
    \foreach \x in {4.5}{\istate{\x}{-8.5}{colSt}}
    \foreach \x in {-3,...,1}{\square{\x+0.5}{\x+1.5}}
	    \foreach \x in {0,1,...,6}{\circle
    {-2.5+\x}{-2.5-\x}}
    \foreach \x in {-3,...,1}{\circle
    {4-\x-0.5}{\x+1.5}}
     \foreach \x in {-10,...,-4}{\circle
    {6.5}{\x+1.5}}
     \circle{3.5}{-8.5}
      \draw[semithick,decorate,decoration={brace}] (-3,-1.25) -- (1.25,3) node[midway,xshift=-8.5pt,yshift=7.5pt] {$\ell-1$};
      \draw[semithick,decorate,decoration={brace}] (2.75,-8.8) -- (-3.5,-2.55) node[midway,xshift=-28pt] {$2t-\ell+1$};
  \draw[semithick,decorate,decoration={brace}] (2.55,2.85) -- (5.8,-0.3) node[midway,xshift=7.5pt,yshift=7.5pt] {$2r$};
  \end{tikzpicture}
\times \,q^{\ell+2r-6t-3}+\begin{tikzpicture}[baseline={([yshift=-4ex]current bounding box.center)},scale=0.45]
{\prop{2}{2}{colSt}{}}
   \foreach \x in {1,3}{\prop{\x}{1}{colSt}{}}   
    \foreach \x in {0,2,...,4}{\prop{\x}{0}{colSt}{}}
    \foreach \x in {-1,1,...,5}{\prop{\x}{-1}{colSt}{}}
    \foreach \x in {-2,0,...,6}{\prop{\x}{-2}{colSt}{}}
    \foreach \x in {-1,1,...,5}{\prop{\x}{-3}{colSt}{}}
        \foreach \x in {1,3,5}{\prop{\x}{-5}{colSt}{}}
                \foreach \x in {3,5}{\prop{\x}{-7}{colSt}{}}
    \foreach \x in {0,2,...,6}{\prop{\x}{-4}{colSt}{}}
        \foreach \x in {2,4,6}{\prop{\x}{-6}{colSt}{}}
                \foreach \x in {4}{\prop{\x}{-8}{colSt}{}}
   \foreach \x in {-8,-6,-4,-2} {\prop{6}{\x}{colimp}{}}
    \foreach \x in {4.5}{\istate{\x}{-8.5}{colSt}}
    \foreach \x in {-3,...,1}{\square{\x+0.5}{\x+1.5}}
	    \foreach \x in {0,1,...,6}{\circle
    {-2.5+\x}{-2.5-\x}}
    \foreach \x in {-3,...,1}{\circle
    {4-\x-0.5}{\x+1.5}}
     \foreach \x in {-10,...,-4}{\circle
    {6.5}{\x+1.5}}
     \circleblack{3.5}{-8.5}
  \end{tikzpicture}
\times \,\frac{(1-p)^{2t-\ell}}{q^{6t+2-\ell-2r}}.
  \ee
By also using the fact that $\begin{tikzpicture}[baseline={([yshift=-0.6ex]current bounding box.center)},scale=0.5]
   \prop{0}{0}{colSt}{}
   \circleblack{-0.5}{-0.5}
    \circleblack{-0.5}{0.5}
 \end{tikzpicture}\ge 
\begin{tikzpicture}[baseline={([yshift=-0.6ex]current bounding box.center)},scale=0.5]
  \circleblack{-0.5}{-0.5}
    \circleblack{-0.5}{0.5}
    \gridLine{-0.38}{-0.5}{0.3}{-0.5}
      \gridLine{-0.38}{0.5}{0.3}{0.5}
 \end{tikzpicture}\,$, we eventually arrive at the result
  \begin{eqnarray}\nonumber
\!\!\!\tr[\rho^2_A(t)]&\geq &\begin{tikzpicture}[baseline={([yshift=-4ex]current bounding box.center)},scale=0.45]
{\prop{2}{2}{colSt}{}}
   \foreach \x in {1,3}{\prop{\x}{1}{colSt}{}}   
    \foreach \x in {0,2,...,4}{\prop{\x}{0}{colSt}{}}
    \foreach \x in {-1,1,...,5}{\prop{\x}{-1}{colSt}{}}
    \foreach \x in {0,2,4,6}{\prop{\x}{-2}{colSt}{}}
    \foreach \x in {1,3,5}{\prop{\x}{-3}{colSt}{}}
        \foreach \x in {3,5}{\prop{\x}{-5}{colSt}{}}
                \foreach \x in {5}{\prop{\x}{-7}{colSt}{}}
    \foreach \x in {2,4,6}{\prop{\x}{-4}{colSt}{}}
        \foreach \x in {4,6}{\prop{\x}{-6}{colSt}{}}
   \foreach \x in {-8,-6,-4,-2} {\prop{6}{\x}{colimp}{}}
    \foreach \x in {-2,...,1}{\square{\x+0.5}{\x+1.5}}
	    \foreach \x in {-1,0,...,6}{\circle
    {-0.5+\x}{-2.5-\x}}
    \foreach \x in {-3,...,1}{\circle
    {4-\x-0.5}{\x+1.5}}
     \foreach \x in {-10,...,-4}{\circle
    {6.5}{\x+1.5}}
    \end{tikzpicture}
\times \frac{1}{q^{3t+2\ell+r}}+\begin{tikzpicture}[baseline={([yshift=-4ex]current bounding box.center)},scale=0.45]
{\prop{2}{2}{colSt}{}}
   \foreach \x in {1,3}{\prop{\x}{1}{colSt}{}}   
    \foreach \x in {0,2,...,4}{\prop{\x}{0}{colSt}{}}
    \foreach \x in {-1,1,...,5}{\prop{\x}{-1}{colSt}{}}
    \foreach \x in {0,2,4,6}{\prop{\x}{-2}{colSt}{}}
    \foreach \x in {1,3,5}{\prop{\x}{-3}{colSt}{}}
        \foreach \x in {3,5}{\prop{\x}{-5}{colSt}{}}
                \foreach \x in {5}{\prop{\x}{-7}{colSt}{}}
    \foreach \x in {2,4,6}{\prop{\x}{-4}{colSt}{}}
        \foreach \x in {4,6}{\prop{\x}{-6}{colSt}{}}
   \foreach \x in {-8,-6,-4,-2} {\prop{6}{\x}{colimp}{}}
    \foreach \x in {-2,...,1}{\square{\x+0.5}{\x+1.5}}
	    \foreach \x in {-1,0,...,6}{\circle
    {-0.5+\x}{-2.5-\x}}
    \foreach \x in {-3,...,1}{\circle
    {4-\x-0.5}{\x+1.5}}
     \foreach \x in {-10,...,-4}{\circle
    {6.5}{\x+1.5}}
     \circleblack{5.5}{-8.5}
               \circleblack{4.5}{-7.5}
               \draw[semithick,decorate,decoration={brace}] (5.25,-8.8) -- (4.,-7.55) node[midway,xshift=-25pt] {$\ell+r-t$};
     \end{tikzpicture}
\times \,\frac{(1-p)^{(2t-\ell)(\ell+r-t)}}{q^{4t+\ell}}\\
&=&\frac{1}{q^{2\ell}}+\begin{tikzpicture}[baseline={([yshift=-4ex]current bounding box.center)},scale=0.45]
{\prop{2}{2}{colSt}{}}
   \foreach \x in {1,3}{\prop{\x}{1}{colSt}{}}   
    \foreach \x in {0,2,...,4}{\prop{\x}{0}{colSt}{}}
    \foreach \x in {-1,1,...,5}{\prop{\x}{-1}{colSt}{}}
    \foreach \x in {0,2,4,6}{\prop{\x}{-2}{colSt}{}}
    \foreach \x in {1,3,5}{\prop{\x}{-3}{colSt}{}}
        \foreach \x in {3,5}{\prop{\x}{-5}{colSt}{}}
                \foreach \x in {5}{\prop{\x}{-7}{colSt}{}}
    \foreach \x in {2,4,6}{\prop{\x}{-4}{colSt}{}}
        \foreach \x in {4,6}{\prop{\x}{-6}{colSt}{}}
   \foreach \x in {-8,-6,-4,-2} {\prop{6}{\x}{colimp}{}}
    \foreach \x in {-2,...,1}{\square{\x+0.5}{\x+1.5}}
	    \foreach \x in {-1,0,...,6}{\circle
    {-0.5+\x}{-2.5-\x}}
    \foreach \x in {-3,...,1}{\circle
    {4-\x-0.5}{\x+1.5}}
     \foreach \x in {-10,...,-4}{\circle
    {6.5}{\x+1.5}}
     \circleblack{5.5}{-8.5}
          \circleblack{4.5}{-7.5}
           \draw[semithick,decorate,decoration={brace}] (5.25,-8.8) -- (4.,-7.55) node[midway,xshift=-25pt] {$\ell+r-t$};
  \end{tikzpicture}
\times \,\frac{(1-p)^{(2t-\ell)(\ell+r-t)}}{q^{4t+\ell}}.
  \end{eqnarray}
By inserting the completeness relation $\1=(\begin{tikzpicture}[baseline={([yshift=-0.6ex]current bounding box.center)},scale=0.65]
   \tgridLine{0}{0}{0.35}{0.35}
  \circle{0}{0}
   \tgridLine{-.5}{0}{-.8}{0.3}
  \circle{-0.5}{0}
\end{tikzpicture}+\begin{tikzpicture}[baseline={([yshift=-0.6ex]current bounding box.center)},scale=0.65]
  \tgridLine{0}{0}{0.3}{0.3}
  \circleblack{0}{0}
  \tgridLine{-.5}{0}{-.8}{0.3}
  \circleblack{-0.5}{0}
  \end{tikzpicture})/q^2$ into legs of impurity gates, we then find that
  \begin{eqnarray}\nonumber
\!\!\!\tr[\rho^2_A(t)]
&\geq&\frac{1}{q^{2\ell}}+\begin{tikzpicture}[baseline={([yshift=-4ex]current bounding box.center)},scale=0.45]
{\prop{2}{2}{colSt}{}}
   \foreach \x in {1,3}{\prop{\x}{1}{colSt}{}}   
    \foreach \x in {0,2,...,4}{\prop{\x}{0}{colSt}{}}
    \foreach \x in {-1,1,...,5}{\prop{\x}{-1}{colSt}{}}
    \foreach \x in {0,2,4,6}{\prop{\x}{-2}{colSt}{}}
    \foreach \x in {1,3,5}{\prop{\x}{-3}{colSt}{}}
        \foreach \x in {3,5}{\prop{\x}{-5}{colSt}{}}
    \foreach \x in {2,4,6}{\prop{\x}{-4}{colSt}{}}
        \foreach \x in {4}{\prop{\x}{-6}{colSt}{}}
   \foreach \x in {-4,-2} {\prop{6}{\x}{colimp}{}}
    \foreach \x in {-2,...,1}{\square{\x+0.5}{\x+1.5}}
	    \foreach \x in {-1,0,...,4}{\circle
    {-0.5+\x}{-2.5-\x}}
    \foreach \x in {-3,...,1}{\circle
    {4-\x-0.5}{\x+1.5}}
     \foreach \x in {-6,...,-4}{\circle
    {6.5}{\x+1.5}}
    \circleblack{5.5}{-5.5}
      \circleblack{4.5}{-6.5}
  \end{tikzpicture}
\times\, \left(\frac{1}{q^2}\,\begin{tikzpicture}[baseline={([yshift=-0.6ex]current bounding box.center)},scale=0.5]
   \prop{1}{0}{colimp}{}
   \circle{1.5}{0.5}
    \circleblack{0.5}{0.5}
       \circle{1.5}{-0.5}
    \circleblack{0.5}{-0.5}
 \end{tikzpicture}\,\right)^{\ell+r-t} \times\,\frac{(1-p)^{(\ell+r-t)(2t-\ell)}}{q^{4t+\ell}}\\&\geq&
\frac{1}{q^{2\ell}}+
\frac{(1-p)^{2(\ell+r-t)(2t-\ell)}}{q^{2\ell}}\times \left(\frac{1}{q^2}\,\begin{tikzpicture}[baseline={([yshift=-0.6ex]current bounding box.center)},scale=0.5]
   \prop{1}{0}{colimp}{}
   \circle{1.5}{0.5}
    \square{0.5}{0.5}
       \circle{1.5}{-0.5}
    \square{0.5}{-0.5}
 \end{tikzpicture}\,-1\right)^{\ell+r-t}, 
  \end{eqnarray} 
where in the last inequality we again used the bound $\begin{tikzpicture}[baseline={([yshift=-0.6ex]current bounding box.center)},scale=0.5]
   \prop{0}{0}{colSt}{}
   \circleblack{-0.5}{-0.5}
    \circle{-0.5}{0.5}
 \end{tikzpicture}\ge(1-p)\, 
\begin{tikzpicture}[baseline={([yshift=-0.6ex]current bounding box.center)},scale=0.5]
  \circle{-0.5}{-0.5}
    \circleblack{-0.5}{0.5}
    \gridLine{-0.38}{-0.5}{0.3}{-0.5}
      \gridLine{-0.38}{0.5}{0.3}{0.5}
 \end{tikzpicture}\,$. This proves the bound given in Eq.~(23) of the main text.


\end{document}